\documentclass[prd,aps,showpacs,superscriptaddress]{revtex4}
\usepackage{epsfig}
\usepackage{color}
\usepackage{amsmath}
\usepackage{amssymb}
\newcommand{\ben}{\begin{eqnarray}}
\newcommand{\een}{\end{eqnarray}}

\newcommand{\bef}{\begin{figure}[htb]\centering}
\newcommand{\eef}{\end{figure}}

\begin{document}

\title{Low-mass lepton pair production 
       at large transverse momentum}

\author{Zhong-Bo Kang} 
\email{kangzb@iastate.edu}
\affiliation{Department of Physics and Astronomy, 
             Iowa State University,  
             Ames, IA 50011, USA }
\author{Jian-Wei Qiu}
\email{jwq@iastate.edu}
\affiliation{Department of Physics and Astronomy, 
             Iowa State University,  
             Ames, IA 50011, USA }
\author{Werner Vogelsang} 
\email{vogelsan@quark.phy.bnl.gov}
\affiliation{Physics Department, Brookhaven National Laboratory,
             Upton, NY 11973, USA }
\date{\today}

\begin{abstract}
We study the transverse momentum distribution of low-mass lepton 
pairs produced in hadronic scattering, using the perturbative QCD 
factorization approach. We argue that the distribution at large transverse 
momentum, $Q_T \gg Q$, with the pair's invariant mass $Q$ as low as 
$Q \sim \Lambda_{\mathrm{QCD}}$, can be systematically factorized 
into universal parton-to-lepton pair fragmentation functions, parton
distributions, and perturbatively calculable partonic hard parts 
evaluated at a short distance scale $\sim {\cal O}(1/Q_T)$.  
We introduce a model for the input lepton pair fragmentation 
functions at a scale $\mu_0\sim 1$~GeV, which are then evolved 
perturbatively to scales relevant at RHIC.  Using the evolved 
fragmentation functions, we calculate the transverse momentum 
distributions in hadron-hadron, hadron-nucleus, and nucleus-nucleus 
collisions at RHIC. We also discuss the sensitivity of the transverse 
momentum distribution of low-mass lepton pairs to the gluon distribution.
\end{abstract}


\maketitle

\section{Introduction}
\label{intro}

Hadronic processes producing a high-transverse-momentum ($Q_T$) 
photon play a fundamental role in High-Energy Physics. They offer 
possibilities to explore QCD, probing for example the structure of the 
interacting hadrons or testing the predictive power of QCD calculations. 
They are intimately involved in many signals (and their backgrounds) for 
New Physics, and may also serve as important probes of the properties 
of the strongly-interacting matter produced in heavy-ion collisions. 
Their ``point-like'' QED coupling to quarks makes photons particularly 
suited for these applications. 

One can distinguish two types of high-$Q_T$ photon signals, depending
on whether the produced photons are real ($Q^2=0$), or virtual 
($Q^2 > 0$). The first case corresponds to the ``classic'' prompt-photon 
situation \cite{prompt:photon}. 
In the second case, the photon could subsequently decay 
into a lepton pair $\ell^+\ell^-$, which can be detected. This
case thus corresponds to a ``Drell-Yan'' type situation, with a
lepton pair at high $Q_T$ 
\cite{Aurenche:1988yr,Berger:1998ev,Qiu:2001nr,Berger:2001wr}. 

Both types have their virtues and disadvantages. Electromagnetic 
probes of QCD hard scattering are generally relatively rare compared
to hadronic ones such as jets. At a given $Q_T$, real photons are much more 
copious than lepton pairs from virtual photon decay, as the latter 
suffer from much lower rates, both due to an additional power of the 
electromagnetic coupling $\alpha_{\mathrm{em}}$ and to a phase space suppression.
On the other hand, one would intuitively expect that virtual photons
could both give experimentally cleaner signals and be theoretically 
more tractable. Real photons may have large backgrounds from $\pi^0$ decay 
or may be faked in the detector by neutral hadrons, in particular toward
lower $Q_T$. While these problems can be addressed in part by applying 
an isolation cut to the photon signal, for lepton pairs they will not be 
there in the first place. On the theoretical side, complications in the 
calculation of the cross section that are present at $Q^2\approx 0$ 
may be alleviated if $Q^2 \gg \Lambda_{\mathrm{QCD}}^2$. These 
complications are related to the 
presence of a ``fragmentation'' (or, ``bremsstrahlung'') contribution to 
the real-photon cross section, which arises from the observed photon being 
produced in jet fragmentation~\cite{prompt:photon}. 
Since this contribution involves 
non-perturbative parton-to-photon fragmentation 
functions~\cite{photff,photff1,photff2}, about which 
there is only limited knowledge, it leads to an uncertainty in the 
theoretical predictions. When, however, $Q$ is much larger than typical 
non-perturbative scales in QCD, $Q_T \gg Q\gg \Lambda_{\mathrm{QCD}}$, 
the fragmentation part can be calculated in factorized perturbation theory, 
as was shown in~\cite{Qiu:2001nr,Berger:2001wr}, which allows rather 
precise theoretical estimates of the cross section. 

Clearly, the situation $Q_T \gg Q\gg \Lambda_{\mathrm{QCD}}$ is ideal 
from a theoretical point of view but tough to realize in experiment, 
due to the low rates of lepton pairs with these kinematics. 
In the present paper, we investigate the photon 
cross section in the regime $Q_T \gg Q \sim \Lambda_{\rm QCD}$. 
Our study is very much motivated by measurements performed for these 
kinematics~\cite{Akiba} by the PHENIX collaboration at the Relativistic 
Heavy Ion Collider (RHIC) at Brookhaven National Laboratory (BNL). 
Choosing lepton pairs with $100 \leq Q \leq 
300$~MeV allows PHENIX to extend their photon measurements to lower 
transverse momenta, $Q_T\leq 5$~GeV, than possible for real prompt 
photons for which the background from neutral hadrons becomes
overwhelming. Thus the improvement in systematic uncertainties offered 
by the low-mass lepton pairs potentially 
outweighs the loss in statistics. At higher 
$Q_T\gtrsim 5$~GeV, it becomes eventually more beneficial to consider real 
photons. However, much of the interesting physics in heavy-ion collisions
or (polarized) $pp$ scattering takes place in the region $Q_T\leq 5$~GeV.

When $Q \sim \Lambda_{\rm QCD}$, the fragmentation contribution
to the cross section will no longer be completely perturbative as it was
for $Q \gg \Lambda_{\rm QCD}$. However, we will argue that, like for 
real photons, the hadronic cross section for producing high transverse 
momentum lepton pairs with low invariant mass
can be systematically factorized into universal fragmentation
functions, parton distributions, 
and perturbatively calculable partonic hard parts evaluated
at a distance scale $\sim {\cal O}(1/Q_T)$. We note that this is similar 
in spirit to what is typically done in the corresponding ``space-like'' 
case of the structure functions of virtual photons~\cite{virtphot}, 
which received a lot of attention at HERA~\cite{bussey}. 
Thus, our main conclusion 
will be that the lepton-pair cross section in this kinematic regime
can be treated in a similar fashion, and with similar rigor, as the 
real-photon one, which makes the efforts to use it experimentally
as a means to access lower $Q_T$ very worthwhile. We will also discuss 
the connection between the very low mass dilepton pair and direct 
real-photon production. We note that there has been
an earlier theoretical study of the dilepton cross section in this
kinematic regime~\cite{Aurenche:1988yr}. Our study goes beyond that
work, particularly by providing a more complete treatment of the
fragmentation contribution and more extensive phenomenological
studies.

We will introduce a model for the input lepton pair fragmentation 
functions at a scale $\mu_0\sim 1$~GeV, which are then evolved perturbatively 
to scales relevant at RHIC. Using the evolved fragmentation functions,
we calculate the transverse momentum distributions of lepton pairs in 
proton-proton collisions at $\sqrt{s}=200$~GeV. We also discuss the 
case of nuclear collisions at RHIC, and investigate the sensitivity of 
high-$Q_T$ low mass dileptons to the nuclear gluon distribution. 

The rest of our paper is organized as follows. In Sec.~\ref{drellyan},
we review the perturbative QCD factorization for the Drell-Yan
cross section when $Q_T\gg Q\gg \Lambda_{\rm QCD}$, to set up the
notation and terminology.  In Sec.~\ref{lowmass}, we study the
asymptotic limit of the factorized cross section as 
$Q\to 0$.  We demonstrate that all perturbative short-distance 
parts of the factorized cross section are finite and can be expanded
as power series in $Q^2/Q_T^2$.  All non-perturbative physics present
when $Q\sim \Lambda_{\rm QCD}$ is systematically included in new
non-perturbative ``parton-to-low-mass-dilepton'' fragmentation 
functions.  We introduce our model for these fragmentation functions
at a scale $\mu_0=1$~GeV.  In Sec.~\ref{number}, we use the 
perturbatively evolved fragmentation functions to obtain 
predictions for $pp$ scattering at RHIC.  We also study
the nuclear dependence of low-mass dilepton transverse momentum
distributions, and their role for extracting nuclear gluon distributions. 
Finally, Section~\ref{summary} presents our summary and conclusions.

\section{Lepton pair production with $Q_T \gg Q \gg \Lambda_{\rm QCD}$}
\label{drellyan}

\begin{figure}[h]
\begin{center}
\psfig{file=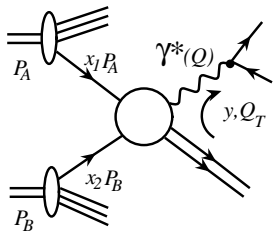, height=1.2in}
\caption{Sketch for the ``Drell-Yan type'' hadronic production of a 
lepton pair via the decay of a virtual photon.}
\label{dilepton}
\end{center}
\end{figure}

The cross section for Drell-Yan type inclusive lepton pair production
in hadronic collisions, 
$A(P_A)+B(P_B)\rightarrow \gamma^*(\rightarrow \ell^+\ell^-(Q))+X$, 
as sketched in Fig.~\ref{dilepton}, 
can be expressed in terms of the cross section for producing a
virtual photon that decays into the observed lepton pair:
\begin{equation}
\frac{d\sigma_{AB\rightarrow \ell^+\ell^-(Q) X}}{dQ^2\,dQ_T^2\,dy}
= \left(\frac{\alpha_{\mathrm{em}}}{3\pi Q^2}\right)
\sqrt{1-\frac{4m_\ell^2}{Q^2}}\left(1+\frac{2m_\ell^2}{Q^2}\right)
  \frac{d\sigma_{AB\rightarrow \gamma^*(Q) X}}{dQ_T^2\,dy}\, ,
\label{DY-Vph}
\end{equation}
where the variables $Q$, $Q_T$, and $y$ are the invariant mass, 
transverse momentum, and rapidity of the virtual photon, respectively.  
$m_\ell$ is the lepton mass, and the symbol $X$ stands 
for an inclusive sum over final states that recoil against 
the virtual photon.  The overall factor on the right-hand-side of 
Eq.~(\ref{DY-Vph}) arises from the integration over the pair's
angular distribution, $\sqrt{1-4m_\ell^2/Q^2}$ representing the mass 
threshold for producing the pair.

When both physically measured momentum scales $Q$ and $Q_T$ 
are much larger than $\Lambda_{\rm QCD}$, 
the cross section for producing the virtual 
photon can be factored systematically in QCD perturbation theory as 
\cite{Collins:1989gx}
\begin{equation}
\frac{d\sigma_{AB\rightarrow \gamma^*(Q) X}}{dQ_T^2\,dy}
=\sum_{a,b}\int dx_1 f_a^A(x_1,\mu) 
           \int dx_2 f_b^B(x_2,\mu)\,
 \frac{d\hat{\sigma}^{\rm Pert}
                    _{ab\rightarrow \gamma^*(Q) X}}{dQ_T^2\,dy}
 (x_1,x_2,Q,Q_T,y;\mu) \, ,
\label{Vph-fac}
\end{equation}
where the sum $\sum_{a,b}$ runs over all parton flavors, 
$x_1$ and $x_2$ are the partons' momentum fractions, and $f_a^A$
and $f_b^B$ are the corresponding parton distributions.
$\mu$ represents the renormalization and factorization scale; 
it is of the order of the energy exchange 
in the reaction: $\mu \sim \sqrt{Q^2 + Q_T^2}$.  The function 
$d\hat{\sigma}^\mathrm{Pert}
              _{ab\rightarrow \gamma^*(Q) X}/dQ_T^2 dy$ 
in Eq.~(\ref{Vph-fac}) represents the short-distance physics of the
collision and is calculable perturbatively in terms of a power
series in $\alpha_s(\mu)$.  

Beyond the leading order in $\alpha_s$, when $Q_T \gg Q$,
the short-distance perturbative functions
$d\hat{\sigma}^{\rm Pert}_{ab\rightarrow \gamma^*(Q) X}/dQ_T^2 dy$ 
in Eq.~(\ref{Vph-fac}) can receive large high order corrections 
in powers of $\alpha_s\ln(Q_T^2/Q^2)$ which are caused by the radiation 
of partons along the direction of the low mass lepton pair.  
Such large logarithmic corrections can be
systematically resummed into the parton-to-virtual photon 
fragmentation functions, $D_{f\to\gamma^*}$ \cite{Qiu:2001nr}.
The perturbative series for the 
$d\hat{\sigma}^{\rm Pert}
              _{ab\rightarrow \gamma^*(Q) X}/dQ_T^2 dy$
can therefore be re-organized into two terms as 
\cite{Berger:2001wr},
\begin{eqnarray}
\frac{d\hat{\sigma}^{\rm Pert}_{ab\rightarrow \gamma^*(Q) X}}
     {dQ_T^2\,dy}(x_1,x_2,Q,Q_T,y;\mu)
&=&
\frac{d\hat{\sigma}^{\rm Dir}_{ab\rightarrow \gamma^*(Q) X}}
     {dQ_T^2\,dy}(x_1,x_2,Q,Q_T,y;\mu,\mu_F)
\nonumber \\  
&+&
\frac{d\hat{\sigma}^{\rm Frag}_{ab\rightarrow \gamma^*(Q) X}}
     {dQ_T^2\,dy}(x_1,x_2,Q,Q_T,y;\mu,\mu_F)\, , 
\label{fac-dir-frag}
\end{eqnarray}
where the superscripts ``Dir'' and ``Frag'' represent the
``direct'' and the ``fragmentation'' contribution, respectively.
The latter includes the perturbative 
fragmentation logarithms and can be further factorized 
as \cite{Qiu:2001nr}, 
\begin{equation}
\frac{d\hat{\sigma}^{\rm Frag}_{ab\rightarrow\gamma^*(Q) X}}
     {dQ_T^2 dy}
= \sum_{c} \int \frac{dz}{z^2}\, 
  \left[
  \frac{d\hat{\sigma}_{ab\rightarrow c X}}
       {dp_{c_T}^2\,dy}\left(x_1,x_2,p_c=\hat{Q}/z;\mu_F\right)
  \right]
  D_{c\rightarrow \gamma^*}(z,\mu_F^2;Q^2) ,
\label{DY-F}
\end{equation}
where the four-momentum $\hat{Q}^\mu$ is defined to be $Q^\mu$ 
at $Q^2=0$, corresponding to the approximation $Q^2\ll Q_T^2$ made
for the fragmentation piece, and where
$d\hat{\sigma}_{ab\rightarrow c X}/dp_{c_T}^2\,dy$ 
is a short-distance hard part for partons of flavors 
$a$ and $b$ to produce a parton of flavor $c$ and 
momentum $p_c=\hat{Q}/z$.  
The fragmentation contribution in Eq.~(\ref{DY-F}) 
shares the typical two-stage generic pattern of the
fragmentation production of a single particle at
large transverse momentum $Q_T$ (e.g., a hadron of mass 
$M_h\ll Q_T$, or a real photon): the production of an on-shell
parton of flavor $c$ at the distance scale $1/Q_T$, convoluted with
a fragmentation function that includes the leading logarithmic
contributions from the ``running'' of the distance scale 
from $1/\mu_F\sim 1/Q_T$ to $1/\mu_{F0}\sim 1/Q$. 
When $Q\gg \Lambda_{\rm QCD}$, the parton-to-virtual photon 
fragmentation function $D_{f\to \gamma^*}$ is perturbative, and so is
the whole resummed fragmentation contribution, 
$\hat{\sigma}^{\rm Frag}$ \cite{Qiu:2001nr}.

The direct contribution in Eq.~(\ref{fac-dir-frag}) is 
perturbatively calculable in a power series of $\alpha_s$. The
coefficient of the term at $n$-th power in $\alpha_s$ is given by 
\cite{Berger:2001wr},
\begin{eqnarray}
&& \frac{d\hat{\sigma}^{{\rm Dir}(n)}_{ab\rightarrow \gamma^*(Q) X}}
     {dQ_T^2\,dy}\left(x_1,x_2,Q,Q_T,y;\mu,\mu_F\right)
\nonumber \\ 
&& {\hskip 0.5in} \equiv 
\frac{d\hat{\sigma}^{{\rm Pert}(n)}_{ab\rightarrow \gamma^*(Q) X}}
     {dQ_T^2\,dy}\left(x_1,x_2,Q,Q_T,y;\mu\right)
-
\frac{d\hat{\sigma}^{{\rm Asym}(n)}_{ab\rightarrow \gamma^*(Q) X}}
     {dQ_T^2\,dy}\left(x_1,x_2,Q,Q_T,y;\mu,\mu_F\right)\,  ,
\label{DY-pdir}
\end{eqnarray}
where $\hat{\sigma}^{{\rm Pert}(n)}$ is the corresponding $n$-th order
coefficient of $\hat{\sigma}^{\rm Pert}$ in Eq.~(\ref{Vph-fac}),
calculated in conventional 
fixed-order QCD perturbation theory. The superscript ``Asym''
refers to an ``asymptotic'' contribution, which is simply the perturbative 
expansion of the fragmentation contribution, 
$\hat{\sigma}^{\rm Frag}$ in Eq.~(\ref{DY-F}) to the same $n$-th
power in $\alpha_s$, {\it including} expansion of the resummed 
fragmentation function. The asymptotic contribution cancels the 
fragmentation logarithms in the perturbative contribution, 
$\hat{\sigma}^{{\rm Pert}(n)}$, order-by-order in powers of $\alpha_s$.  
Therefore, the direct contribution is free of 
the large fragmentation logarithms, while it still 
keeps some information about the physics between the scales
$Q_T$ and $Q$ from non-logarithmic terms, such as powers of $Q^2/Q_T^2$.
These non-logarithmic terms in the direct contribution enable the 
new re-organized factorization formalism to cover a wide
range of kinematics and to have a smooth transition from the 
low $Q_T$ to the high $Q_T$ region.

The separation between the direct and the fragmentation contribution 
in Eq.~(\ref{fac-dir-frag}) depends on the definition of
the parton-to-virtual photon fragmentation functions.  
A definition or a scheme choice for the fragmentation functions 
uniquely fixes the asymptotic contribution, which then fixes 
the direct contribution, in such a way as to render the sum 
$\hat{\sigma}^{\rm Pert}$ independent of the adopted scheme.

The resummation of the fragmentation logarithms into the
parton-to-virtual photon fragmentation functions is achieved by solving
the inhomogeneous evolution equations \cite{Qiu:2001nr}, 
\begin{eqnarray}
\mu_F^2 \frac{d}{d\mu_F^2} D_{c\rightarrow\gamma^*}(z,\mu_F^2;Q^2)
&=& 
\left(\frac{\alpha_{\mathrm{em}}}{2\pi}\right)
 \gamma_{c\rightarrow\gamma^*}(z,\mu_F^2,\alpha_s;Q^2)
\nonumber \\
&+&
\left(\frac{\alpha_{s}}{2\pi}\right)
\sum_d \int_z^1 \frac{dz'}{z'}
P_{c\rightarrow d}(\frac{z}{z'},\alpha_s)\, 
 D_{d\rightarrow\gamma^*}(z',\mu_F^2;Q^2)\, ,
\label{RG-unpol}
\end{eqnarray}
where $c,d=q,\bar{q},g$.  The ambiguity in defining the 
fragmentation function is connected to the choice of the fragmentation
scale, $\mu_F$, and the removal of the perturbative ultraviolet 
divergence in deriving the evolution kernels.  
In Eq.~(\ref{RG-unpol}), the evolution kernels 
$P_{c\rightarrow d}$ are evaluated at a single hard scale,
$\mu_F$, and can be calculated perturbatively as a power series in 
$\alpha_s$.  The inhomogeneous term in the evolution equations can
also be calculated perturbatively, and in general it has
power correction terms of the form $Q^2/\mu_F^2$, due to the 
virtuality of the photon $Q^2$. In the invariant mass cut-off
scheme \cite{Qiu:2001nr}, the lowest order quark-to-virtual 
photon evolution kernel is given by, 
\begin{equation}
\gamma_{q\rightarrow \gamma^*}^{(0)}(z,k^2;Q^2) 
=
e_q^2 \Bigg[\frac{1+(1-z)^2}{z}
           -z\left(\frac{Q^2}{zk^2}\right) \Bigg]
      \theta(k^2-\frac{Q^2}{z})\, ,
\label{Gq0-unpol}
\end{equation}
where $k^2$ is the invariant mass of the parent quark and 
is identified as $\mu_F^2$, and where the $\theta$-function is 
a consequence of the mass threshold. 
The gluon-to-virtual-photon evolution kernel vanishes at the lowest order 
\begin{equation}
\gamma_{g\rightarrow \gamma^*}^{(0)}(z,k^2;Q^2) 
= 0\, ,
\label{Gg0-unpol}
\end{equation}
because the gluon does not interact directly with the virtual photon. 
The choice of factorization scheme is not unique.  But, some choices,
such as the modified minimum subtraction ($\overline{\rm MS}$) scheme,  
may not respect the mass threshold when $Q^2\neq 0$ and 
lead to negative fragmentation functions 
\cite{Qiu:2001nr,Braaten:2001sz}. 
QCD corrections to the lowest order parton-to-virtual-photon  
splitting function $\gamma_{c\rightarrow\gamma^*}^{(0)}$ can be evaluated 
in principle order-by-order in $\alpha_s$.

If $Q\gg \Lambda_{\rm QCD}$,
the parton-to-virtual photon fragmentation functions are completely 
perturbative.  The input distributions for the evolution
equation in Eq.~(\ref{RG-unpol}) are given by 
\begin{equation}
D_{c\rightarrow \gamma^*}(z,\mu_F^2\le Q^2/z;Q^2) = 0\, ,
\label{pert_input}
\end{equation}
for any flavor $c$, if we choose the invariant mass cut-off scheme for
the fragmentation functions \cite{Qiu:2001nr,Berger:2001wr}.

We conclude this section by summarizing the above equations and 
writing down a compact expression for the hadronic dilepton production 
cross section at high transverse momentum:
\begin{eqnarray}
\frac{d\sigma_{AB\rightarrow \ell^+\ell^-(Q) X}}{dQ^2\,dQ_T^2\,dy}
&=& \left(\frac{\alpha_{\mathrm{em}}}{3\pi Q^2}\right)
\sqrt{1-\frac{4m_\ell^2}{Q^2}}\left(1+\frac{2m_\ell^2}{Q^2}\right)
\nonumber\\
&\times &
\left[ \sum_{a,b} f_a^A\otimes f_b^B\otimes
 \frac{d\hat{\sigma}^{\rm Dir}_{ab\rightarrow \gamma^*(Q) X}}{dQ_T^2\,dy} 
+\sum_{a,b,c} f_a^A \otimes f_b^B \otimes
 \frac{d\hat{\sigma}_{ab\rightarrow c X}}{dp_{c_T}^2\,dy} \otimes
 D_{c\rightarrow \gamma^*} \right] \, ,
\label{mastereq}
\end{eqnarray}
where $\otimes$ represents the convolution over parton momentum 
fractions.   We stress that when both $Q_T$ and $Q$
are perturbative scales, the factorization into the direct
and fragmentation contribution is merely a re-organization of
$\hat{\sigma}^{\rm Pert}$ with the fragmentation logarithms resummed 
to all orders.  The re-organized perturbative expansion 
in Eq.~(\ref{mastereq}) is perturbatively reliable for the full 
kinematic regime $Q_T\sim Q$ to $Q_T\gg Q$.

\section{Lepton pair production at $Q_T \gg Q 
\gtrsim\Lambda_{\rm QCD}$}
\label{lowmass}

In this section, we examine the validity of the factorized formula 
in Eq.~(\ref{mastereq}) when the invariant mass of the lepton pair 
decreases to $Q\gtrsim \Lambda_{\rm QCD}$. 

The short-distance hard parts for the direct and the fragmentation 
contribution, 
$d\hat{\sigma}^{\rm Dir}_{ab\rightarrow \gamma^*(Q) X}/dQ_T^2\,dy$
and $d\hat{\sigma}_{ab\rightarrow c X}/dp_{c_T}^2\,dy$ in 
Eq.~(\ref{mastereq}), are evaluated at the distance scales
$\sim {\cal O}(1/Q_T)$ and $\sim {\cal O}(1/p_{c_T})
\sim{\cal O}(1/Q_T)$, respectively.
Both hard parts are not sensitive to the dynamics at the scale
$Q$ and are infrared safe perturbatively, even as $Q\to 0$.

\begin{figure}[h]
\begin{center}
\psfig{file=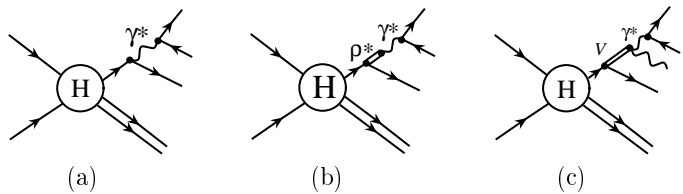, height=1.0in}
\caption{Partonic scattering amplitudes for producing a large $Q_T$
  lepton pair of very low invariant mass $Q$ by (a) a QED
  interaction via a virtual photon, (b) hadronic
  production of a virtual $\rho$ meson which then fluctuates into 
  a virtual photon, or (c) a vector boson decay into a virtual photon 
plus unobserved final state.}
\label{fig1}
\end{center}
\end{figure}

\begin{figure}[h]
\begin{center}
\psfig{file=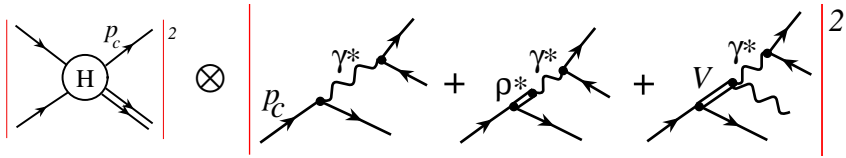, height=0.7in}
\caption{Sketch for the leading-power factorized fragmentation
  contribution to the production of low-mass lepton pairs at large
  transverse momentum.} 
\label{fig2}
\end{center}
\end{figure}

However, when $Q$ decreases to 1 GeV and below, there are 
significant non-perturbative sources to produce the low mass 
lepton pairs.  Other than the Drell-Yan mechanism, or the QED 
production via a virtual photon shown in Fig.~\ref{fig1}(a), 
lepton pairs can also be produced by the conversion of a virtual 
intermediate $\rho$ meson, as shown in Fig.~\ref{fig1}(b), 
as well as in the decay of a light mass vector meson $V$,
see Fig.~\ref{fig1}(c).  
Since the vector meson is unlikely to be produced at the
distance scale of ${\cal O}(1/Q_T)$, these new channels of
dilepton production are only relevant to the fragmentation
contribution, and more precisely, to the fragmentation functions.
That is, the fragmentation contribution to the production of 
very low-mass lepton pairs can still be factorized  
in the same way as that for producing a real photon or 
a single hadron at large transverse momentum, as 
indicated in Fig.~\ref{fig2}.  Therefore,
we expect the QCD factorized formalism for low-mass dilepton 
production at high transverse momentum in Eq.~(\ref{mastereq})
to be valid even when the invariant mass of the lepton pairs is 
as low as $100 \leq Q \leq 300$~MeV, as considered at RHIC \cite{Akiba}.  
When the dilepton's invariant mass $Q$ is of the order of 
$\Lambda_{\rm QCD}$, the only change to Eq.~(\ref{mastereq}) is 
that the parton-to-virtual photon fragmentation function, 
$D_{c\to\gamma^*}$, is no longer fully perturbative. 
The predictive power of the factorized formalism 
in Eq.~(\ref{mastereq}) will rely on the universality of the 
parton-to-virtual photon fragmentation functions.

The parton-to-virtual photon (or lepton pair) fragmentation 
functions are defined in terms of matrix elements of gauge invariant 
non-local operators \cite{Qiu:2001nr}. 
The resummation of the perturbative fragmentation logarithms into 
the fragmentation functions is still achieved by 
solving the same inhomogeneous evolution equations 
as in Eq.~(\ref{RG-unpol}), except that 
the ``natural'' boundary condition in Eq.~(\ref{pert_input}) 
needs to be replaced by non-perturbative input fragmentation 
functions at an initial scale, say, $\mu_0\sim 1$~GeV.  
The inhomogeneous term in the evolution equations is still 
needed because of the perturbative QED production of the lepton
pair via a virtual photon, as shown in Fig.~\ref{fig1}(a).

In principle, the input fragmentation functions
should be universal and determined by experimental data, 
like the fragmentation functions for producing a real photon 
or a single hadron.  In order to estimate the size of 
the production rate of low mass lepton pairs
at high transverse momentum, we propose a model for the input 
fragmentation functions with a minimal number of parameters,
which we hope will serve as an initial step in determining the 
non-perturbative fragmentation functions.

The ``natural'' boundary condition in Eq.~(\ref{pert_input}) is 
a consequence of the pure Drell-Yan production of lepton pairs via 
a single intermediate virtual photon, which does not involve any 
intermediate hadronic meson states, as shown in Fig.~\ref{fig1}(a).
With the additional non-perturbative sources of lepton pair 
production, like those shown in Fig.~\ref{fig1}(b) and (c), 
we assume the following form of the non-perturbative input 
fragmentation functions at the scale $\mu_0= 1$~GeV:
\begin{equation}
D_{f\rightarrow \ell^+\ell^-(Q)}(z,\mu_0^2;Q^2)
=\left(\frac{\alpha_{\mathrm{em}}}{3\pi Q^2}\right)
\sqrt{1-\frac{4m_\ell^2}{Q^2}}\left(1+\frac{2m_\ell^2}{Q^2}\right)
D_{f\rightarrow \gamma^*}(z,\mu_0^2;Q^2)\, ,
\label{frag_input}
\end{equation}
where the parton-to-virtual photon part is given by a sum
of the ``QED'' contribution and a non-perturbative part:
\begin{eqnarray}
D_{f\rightarrow \gamma^*}(z,\mu_0^2;Q^2)
\equiv
D_{f\rightarrow \gamma^*}^{\rm QED}(z,\mu_0^2;Q^2)
+D_{f\rightarrow \gamma^*}^{\rm NonPert}(z,\mu_0^2;Q^2)
\label{full_input}
\end{eqnarray}
($f=q,\bar{q}, g$).  
In the invariant mass cut-off scheme \cite{Qiu:2001nr}, 
we propose the lowest order QED contribution as
\begin{eqnarray}
D_{q\rightarrow\gamma^*}^{\rm QED(0)}(z,\mu_0^2;Q^2) 
& = &
e_q^2 \left( \frac{\alpha_{em}}{2\pi} \right) 
\left[\left( \frac{1+(1-z)^2}{z} \right) 
      \ln\left(\frac{\mu_0^2}
                    {Q^2/z+\lambda^2} \right)
-\left(\frac{Q^2}{Q^2/z+\lambda^2} - 
\frac{Q^2}{\mu_0^2} \right)
\right]\, ,
\nonumber \\
D_{\bar{q}\rightarrow\gamma^*}^{\rm QED(0)}(z,\mu_0^2;Q^2) 
&=&
D_{q\rightarrow\gamma^*}^{\rm QED(0)}(z,\mu_0^2;Q^2) \, ,
\nonumber \\
D_{g\rightarrow\gamma^*}^{\rm QED(0)}(z,\mu_0^2;Q^2) 
&=& 
0\, .
\label{Dq0}
\end{eqnarray}
In deriving the above lowest-order QED fragmentation functions,
we have integrated the invariant mass squared of the fragmenting quark, 
$k^2$, from its lowest value needed to produce a lepton pair of 
invariant mass $Q$ to the factorization scale $\mu_0^2$.   When 
$Q$ is small, the lower limit of $k^2$ could become sensitive to 
the current quark mass and to non-perturbative strong interaction
dynamics at the scale of $\Lambda_{\rm QCD}$. We introduce a mass 
parameter $\lambda$ to prevent the $k^2$ integration from 
entering this strongly interacting regime.  
If one calculated the lowest-order QED quark-to-virtual-photon 
fragmentation function with an effective quark mass, $m_q$, 
one would find that $\lambda^2=m_q^2\, z/(1-z)$ and $z<1$ 
for a non-vanishing $m_q$.  However, the value of 
$\lambda^2$ cannot be derived perturbatively.  It should be 
much less than $\mu_0^2 \sim $1~GeV$^2$ and should be fixed 
by experimental data.  

When $\lambda^2=0$ or $Q^2/z\gg \lambda^2$, 
the lowest order QED fragmentation functions 
in Eq.~(\ref{Dq0}) are reduced to those derived in 
Ref.~\cite{Qiu:2001nr}.  In order 
to maintain the positivity of the lowest order fragmentation function,
and to satisfy the kinematic constraint in Eq.~(\ref{Gq0-unpol}), the
perturbative input distribution in Eq.~(\ref{Dq0}) is valid for 
$z\ge Q^2/(\mu_0^2-\lambda^2) \sim Q^2/\mu_0^2$, which is satisfied 
for the kinematics of interest at RHIC. 

The non-perturbative part of the input fragmentation functions in
Eq.~(\ref{full_input}) should cover all other possible contributions 
including those in Fig.~\ref{fig1}(b) and (c) and their potential 
interferences with each other and 
that in Fig.~\ref{fig1}(a). Similar to what is done in the
case of photon structure functions~\cite{virtphot} 
or fragmentation functions~\cite{photff,photff1}, 
we estimate these 
contributions by adopting the vector meson dominance (VMD) model,
for which
\begin{equation}
\frac{D^{\rm NonPert}_{f\to \gamma}(z,\mu_0^2;Q^2)}
     {D_{f\to V}(z,\mu_0^2)}
\propto \frac{e^2}{f_V^2}=\frac{4\pi\alpha_{\mathrm{em}}}{f_V^2}\, ,
\label{vmd}
\end{equation}
with the fragmentation function $D_{f\to V}$ for a parton of flavor $f$ to 
a vector meson $V$, and the vector meson decay constant 
$f_V$.  Following the Kroll-Wada formula for the Dalitz decay of vector
mesons \cite{Kroll:1955zu}, we multiply the fragmentation function 
in Eq.~(\ref{vmd}) by a form factor and a phase-space 
correction factor, to obtain
\begin{equation}
D^{\rm NonPert}_{q\to\gamma^*}(z,\mu_0^2;Q^2)
\propto D_{q\to V}(z,\mu_0^2)\,
\frac{4\pi\alpha_{\mathrm{em}}}{f_V^2}\,
|F(Q^2)|^2
\left(1-\frac{Q^2}{m_{V}^2}\right)^3 \, ,
\label{vmd-m}
\end{equation}
with $F(Q^2=0)=1$ \cite{Kroll:1955zu,Faessler:1999de}. 
In order to take into account any other possible contributions and
the potential interference between different production amplitudes, we 
introduce an unknown constant $\kappa \sim {\cal O}(1)$, and choose
\begin{equation}
D^{\rm NonPert}_{q\to\gamma^*}(z,\mu_0^2;Q^2)
\equiv \kappa\, D_{q\to V}(z,\mu_0^2)\,
\frac{4\pi\alpha_{\mathrm{em}}}{f_V^2}\,
\left(1-\frac{Q^2}{m_{V}^2}\right)^3 \, .
\label{Dnonpert}
\end{equation}
Here we have absorbed 
the unknown form factor $F(Q^2)$ into the parameter $\kappa$, in order 
to minimize the number of free parameters.  
Although the form factor $F(Q^2)$ depends on $Q^2$, we expect
a weak $Q^2$ dependence when $Q^2$ is small \cite{Faessler:1999de}. 
Following~\cite{photff,photff1},
we further assume that $m_V=m_\rho$, $f_\rho^2/4\pi=2.2$, and 
$D_{f\to V}\approx D_{f\to\pi}$.
For our numerical calculations in Sec.~\ref{number}, we thus choose
the input fragmentation functions as
\begin{equation}
D_{f\to \gamma^*}(z,\mu_0^2;Q^2)
=
D^{\rm QED(0)}_{f\to\gamma^*}(z,\mu_0^2;Q^2)
+\kappa\, D_{f\to \pi}(z,\mu_0^2)\,
\frac{4\pi\alpha_{\mathrm{em}}}{f_\rho^2}\,
\left(1-\frac{Q^2}{m_{\rho}^2}\right)^3 \, .
\label{full_frag}
\end{equation}
We note that future precise experimental data may well require
a more refined ansatz, for which one could make use of the details
for vector meson form factors and decay rates given in \cite{Faessler:1999de}.

Note that we have made a ``one-photon'' approximation throughout, 
meaning that the lepton pair is always eventually produced by the decay
of a single virtual photon. 
In this approximation, 
the low-mass dilepton production 
cross section is equal to the cross section for  
producing a virtual photon at large transverse momentum,
multiplied by a calculable QED factor for the virtual
photon to decay into the measured lepton pair, as expressed
in Eq.~(\ref{DY-Vph}).  
Other than the explicit $Q^2$ dependence in the 
overall QED decay factor, the perturbatively factorized
formula in Eq.~(\ref{mastereq}) indicates that 
any remaining $Q^2$ dependence 
can only come from the short-distance part of the 
direct contribution, 
$d\hat{\sigma}^{\rm Dir}_{ab\rightarrow \gamma^*(Q) X}/dQ_T^2\,dy$, 
and from the non-perturbative parton-to-virtual-photon fragmentation 
functions, $D_{c\to\gamma^*}$.

Like for the real photon cross section \cite{prompt:photon}, 
the direct contribution in Eq.~(\ref{mastereq}) is only 
sensitive to the dynamics at a distance scale $1/Q_T \sim 1/\mu_F$.
The logarithmic infrared sensitivity near $Q\gtrsim \Lambda_{\rm QCD}$
from the Feynman diagrams at next-to-leading order (NLO) or
higher orders is canceled between the two terms 
in Eq.~(\ref{DY-pdir}) order-by-order in  $\alpha_s$.  
The direct contribution is finite when $Q^2/Q_T^2\to 0$, and  
its $Q$-dependence could be expanded in terms of a power series in 
$Q^2/Q_T^2 \sim Q^2/\mu_F^2$.

The leading order (LO) direct contribution to virtual photon 
production is independent of the factorization scheme adopted and 
becomes the same as the LO direct contribution to the production of a
real photon when $Q^2/Q_T^2\to 0$.  However, at higher orders in
$\alpha_s$, the direct contributions to virtual or real photon
production depend on the factorization scheme and do not necessarily
have to be the same when $Q^2/Q_T^2\to 0$.
For massless prompt photon production, the short-distance
hard parts are commonly evaluated by using dimensional 
regularization and the $\overline{\rm MS}$ factorization scheme 
\cite{prompt:photon}. This means that also collinear singularities
in the final state, when a final-state 
quark radiates the real photon, are subtracted 
in $\overline{\rm MS}$. In contrast to this, in our framework for the direct 
contributions to virtual photon production
(see Eq.~(\ref{DY-pdir})), we subtract any collinear singularities
that arise for $Q^2/Q_T^2\to 0$ through the asymptotic term, which
is a procedure different from $\overline{\rm MS}$ subtraction.
Therefore, starting from NLO, the 
direct contribution to virtual photon production in the limit 
$Q^2/Q_T^2\to 0$ differs from that for real photon production.
But, 
the total production cross section, i.e. the sum of the direct and 
fragmentation contributions, is not sensitive to the 
choice of factorization scheme \cite{Qiu:2001nr,Berger:2001wr}.   

Partons produced in hard collisions could fragment into virtual 
photons with either transverse or longitudinal polarization
when the photon's invariant mass $Q$ is finite.   
The fragmentation function for producing longitudinally polarized 
virtual photons is only significant when the value of the effective 
momentum fraction $z$ is very small \cite{Qiu:2001nr,Braaten:2001sz}.
Because of the steep fall-off of the parton distribution functions 
as the parton momentum fraction $x$ increases, the virtual photon 
cross section is dominated by a relatively large $z$ 
in the fragmentation functions \cite{Berger:2001wr}, and thus, 
most large-$Q_T$ virtual photons produced in 
hadronic collisions at RHIC energies should be transversely 
polarized, like the real photon \cite{Qiu:2001ac}.

As the invariant mass $Q$ of the lepton pair decreases,
we expect the virtual photon cross section to become close to
the one for producing a real photon, except however for the 
(approximately) logarithmic $Q^2$ dependence of the parton-to-virtual photon
fragmentation functions caused by the radiation of a
low mass virtual photon which decays into the measured lepton 
pair.  The same QED bremsstrahlung for producing a real 
photon is absorbed into the parton-to-real-photon 
fragmentation functions, 
which of course do not have any explicit $Q^2$ 
dependence and hence no logarithm in the input distribution.  
On the other hand, the 
QED channel to produce a lepton pair via a virtual photon,
like the one in Fig.~\ref{fig2}(a), can be perturbatively 
calculated, and the potential logarithmic divergence as $Q$ 
decreases is naturally cut off by the threshold factor 
$\sqrt{1-4m_\ell^2/Q^2}$ in Eq.~(\ref{mastereq}) and/or the 
non-perturbative mass parameter $\lambda^2$ in Eq.~(\ref{Dq0}).  
That is, the logarithmic $Q^2$ 
dependence of parton-to-virtual-photon fragmentation functions
from the lowest order QED calculation in Eq.~(\ref{Dq0}) 
(or, the improved $Q^2$ dependence after including higher order 
contributions) is physical.  
The $Q^2$ dependence is measurable via the 
invariant mass of the measured lepton pair.  We will discuss 
the consequence of this difference in the $Q^2$ dependence between 
the virtual and real photon cross sections in the next section.

\section{Numerical results}
\label{number}

In this section, we discuss the phenomenology of low-mass
dilepton production in proton-proton as well as nuclear collisions 
at RHIC. We will mostly consider electron-positron pairs, 
as relevant for the PHENIX measurements at RHIC~\cite{Akiba}.

\subsection{Low-mass lepton pairs at large transverse momentum in $pp$
collisions} 

We define the invariant cross section for producing lepton pairs 
in a mass range between $Q_{\rm min}$ and $Q_{\rm max}$ as 
\begin{eqnarray}
E\frac{d\sigma_{AB\to\ell^+\ell^-(Q)X}}{d^3Q}
\equiv 
\int_{Q_{\rm min}^2}^{Q_{\rm max}^2} dQ^2 \, \frac{1}{\pi} \,
\frac{d\sigma_{AB\rightarrow \ell^+\ell^-(Q) X}}
{dQ^2\,dQ_T^2\,dy} \, ,
\label{inv_xsec}
\end{eqnarray}
where the differential dilepton cross section is given 
in Eq.~(\ref{mastereq}).  

The low-mass dilepton cross section 
vanishes at $Q=2m_\ell$ due to the threshold factor
$\sqrt{1-4m_\ell^2/Q^2}$ in Eq.~(\ref{mastereq}). This value of $Q$ is 
far lower than the values $Q_{\rm min}=100$~MeV and $Q_{\rm max}=300$~MeV
we will consider here. Therefore, among the kinematic factors in 
Eq.~(\ref{mastereq}) only the term $1/Q^2$ is relevant in the integration
over $Q^2$. We note that in fact for $Q\gg 2m_\ell$ one
has $\sqrt{1-4m_\ell^2/Q^2}\left(1+2m_\ell^2/Q^2\right)
=1+{\cal O}(m_\ell^4/Q^4)$ in~(\ref{mastereq}), 
amounting to only a tiny correction
in the mass region we are considering. Furthermore,
as we discussed in the previous section, at $Q\ll Q_T$ the factorized 
perturbative cross section for producing a virtual photon 
becomes identical to the prompt photon one, except for the approximately
logarithmic $Q^2$ dependence in the QED parton-to-virtual-photon 
fragmentation functions. Therefore, 
we could estimate the invariant cross section in Eq.~(\ref{inv_xsec}) 
as follows:
\begin{eqnarray}
E\frac{d\sigma_{AB\to\ell^+\ell^-(Q)X}}{d^3Q}
&\approx &
\frac{d\sigma_{AB\rightarrow \gamma(\hat{Q}) X}}{dQ_T^2\,dy}\,
\int_{Q_{\rm min}^2}^{Q_{\rm max}^2} dQ^2 \, 
\left(\frac{\alpha_{\mathrm{em}}}{3\pi^2 Q^2}\right)
\sqrt{1-\frac{4m_\ell^2}{Q^2}}\left(1+\frac{2m_\ell^2}{Q^2}\right)
\nonumber\\
&\approx &
\frac{\alpha_{\mathrm{em}}}{3\pi}\, 
\ln\left(\frac{Q_{\rm max}^2}{Q_{\rm min}^2}\right)\,
E_{\gamma}\frac{d\sigma_{AB\to\gamma(\hat{Q})X}}{d^3Q}
\label{photon} \, ,
\end{eqnarray}
where $\hat{Q}^2=0$, and
$E_{\gamma}d\sigma_{AB\to\gamma(\hat{Q})X}/d^3Q$ is 
the prompt photon cross section.  

Eq.~(\ref{photon}) should be a good approximation if 
the virtual photon cross section has a very mild $Q^2$ dependence 
and is close to the prompt photon one
when $Q^2_{\rm min} \leq Q^2 \leq Q^2_{\rm max}$.  
To test this real photon approximation, we introduce an
invariant virtual photon cross section,
\begin{equation}
E\frac{d\sigma_{AB\rightarrow \gamma^*(Q) X}}{d^3Q}
=\frac{1}{\pi}
\frac{d\sigma_{AB\rightarrow \gamma^*(Q) X}}{dQ_T^2\,dy}
\approx
\left(\frac{3Q^2}{\alpha_{\mathrm{em}}}\right)
\frac{d\sigma_{AB\to\ell^+\ell^-(Q)X}}{dQ^2dQ_T^2dy}\, ,
\label{vphoton}
\end{equation}
where the dilepton differential cross section is given
in Eq.~(\ref{mastereq}) and terms of order $m_{\ell}^2/Q^2$ 
have been neglected. 

In order to generate the numerical values for the dilepton 
cross section in the figures below, we let 
the fragmentation scale, $\mu_F$, be equal
to the renormalization scale, $\mu$, and choose 
$\mu_F=\mu = \sqrt{Q^2+Q_T^2}$.  We use the NLO CTEQ6M parton
distributions for the unpolarized nucleon \cite{CTEQ6}. 
We solve for the parton-to-virtual photon fragmentation functions 
by using the inhomogeneous evolution equation
in Eq.~(\ref{RG-unpol}) and the input fragmentation
functions in Eq.~(\ref{full_frag}) at $\mu_0=1$~GeV and 
$\lambda=200$~MeV. 
For the parton-to-pion fragmentation functions needed for 
the non-perturbative input fragmentation functions 
in Eq.~(\ref{full_frag}), we use the parameterization derived in
Ref.~\cite{KKP}.

\begin{figure}[h]
\begin{center}
\psfig{file=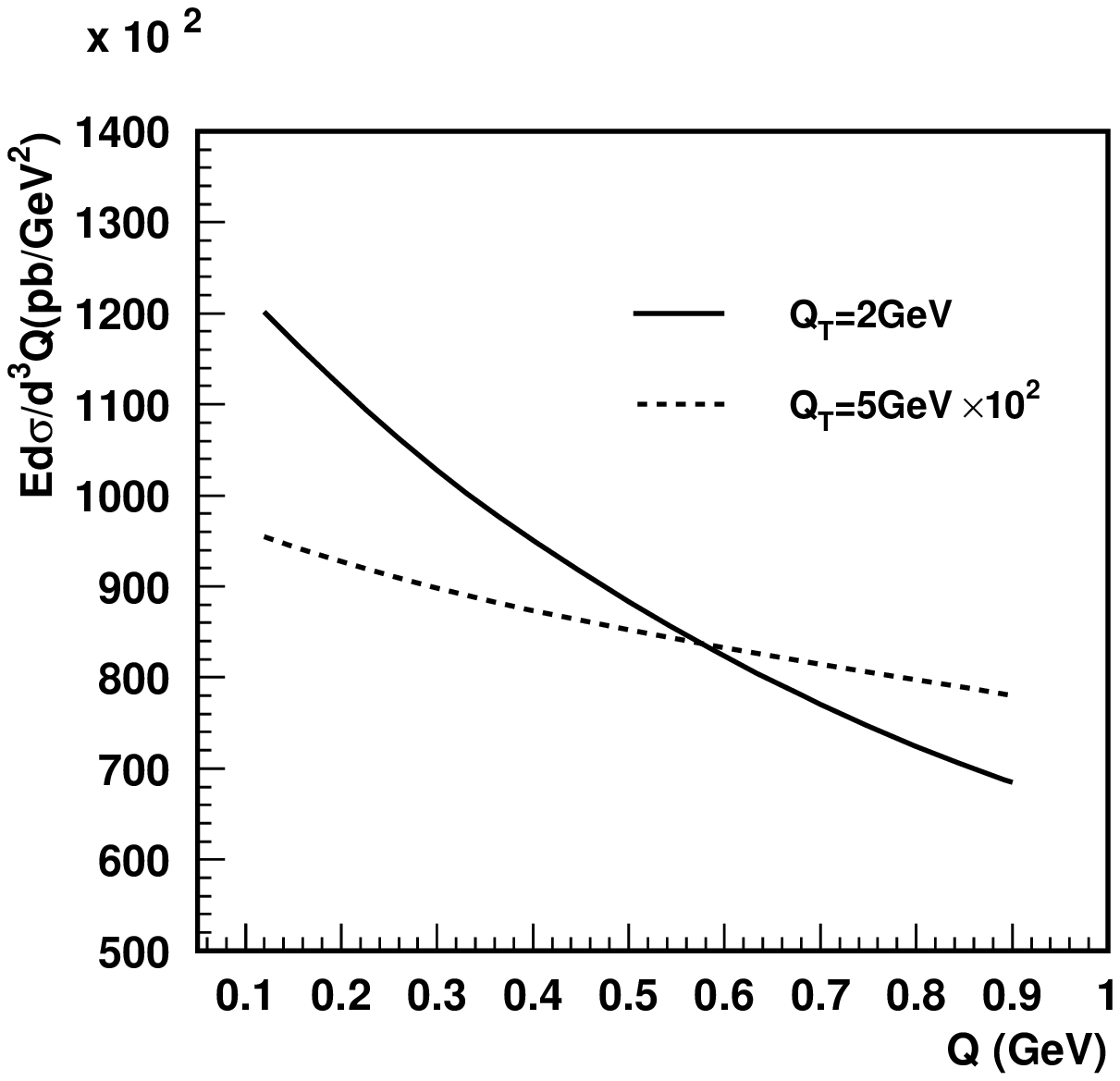,width=3.0in}
\hskip 0.2in
\psfig{file=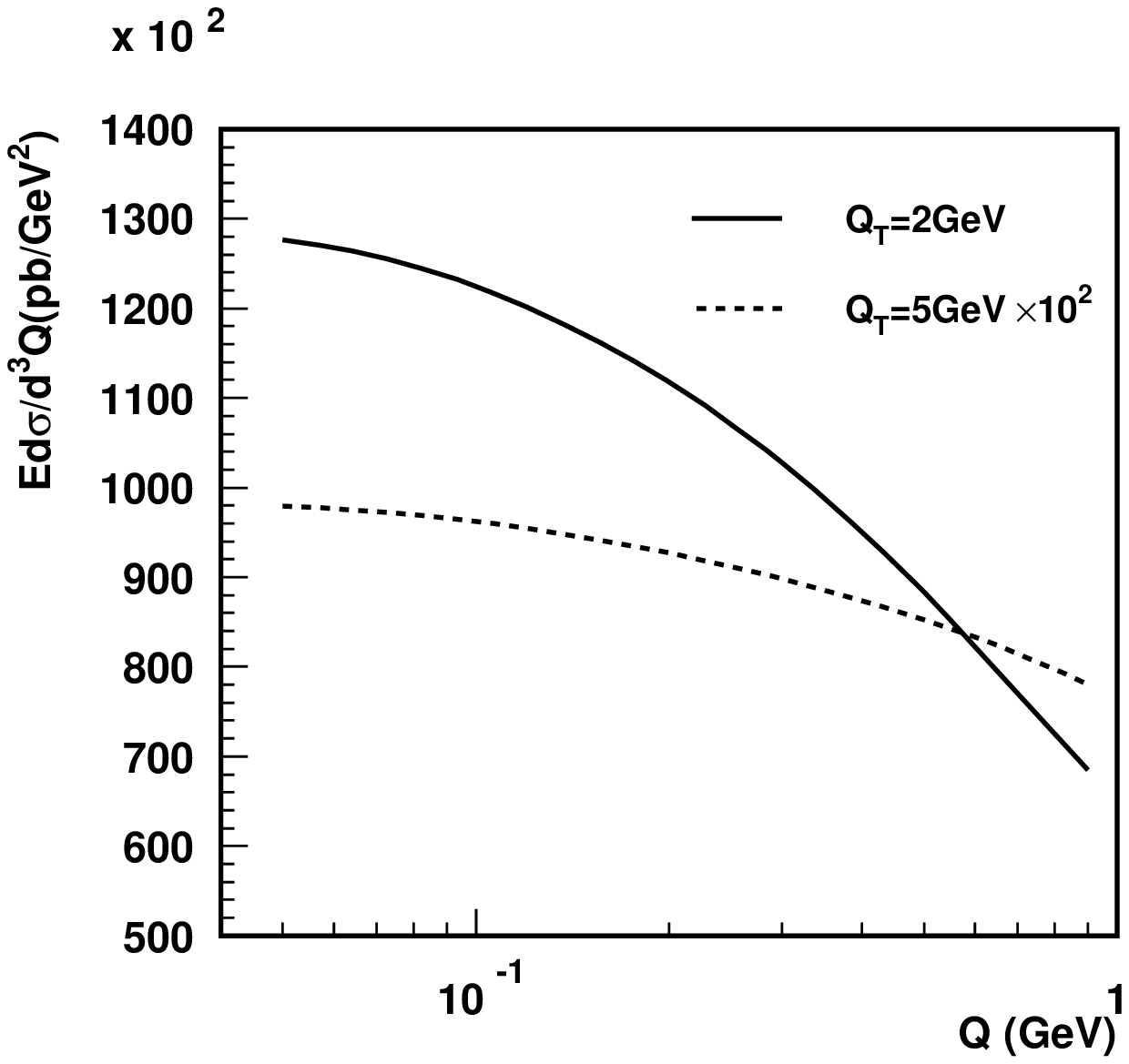,width=3.0in}
\caption{Invariant virtual photon cross section 
in Eq.~(\ref{vphoton}) as a function of the 
photon's invariant mass $Q$ in proton-proton collision 
at $\sqrt{s}=200$~GeV, $y=0$, and $Q_T=2,\, 5$ GeV. Note that
the results for $Q_T=5$ GeV have been multiplied by a factor 100.
The plot on the right-hand-side has a logarithmic scale of the 
abscissa, in order to better illustrate the small-$Q$ region.}
\label{fig4}
\end{center}
\end{figure}

In Fig.~\ref{fig4}, we plot the invariant virtual photon 
cross section as a function of $Q$ in proton-proton collisions
at $\sqrt{s}=200$~GeV and rapidity $y=0$.  
The solid line is for $Q_T$=2~GeV, while 
the dashed one is for $Q_T$=5~GeV.  
From the figure on the left it is clear that the $Q^2$ dependence 
of the virtual photon cross section is relatively weak
for $Q_T$=5~GeV, while it becomes more significant when $Q_T$ 
is small. Therefore, for the range of $Q$ of interest for PHENIX 
\cite{Akiba}, 
the real photon approximation in Eq.~(\ref{photon}) 
should be reasonable if $Q_T > 5$~GeV, 
while it will become less accurate below that.  In the figure
on the right, we plot the cross section for the $Q$ as small as
50~MeV.  The logarithmic plot clearly shows the slowing down 
of the growth of the cross section 
when $Q\lesssim \lambda = 200$~MeV. The virtual photon cross
section should asymptotically approach the real photon 
cross section as $Q\to 0$.

\begin{figure}[h]
\begin{center}
\psfig{file=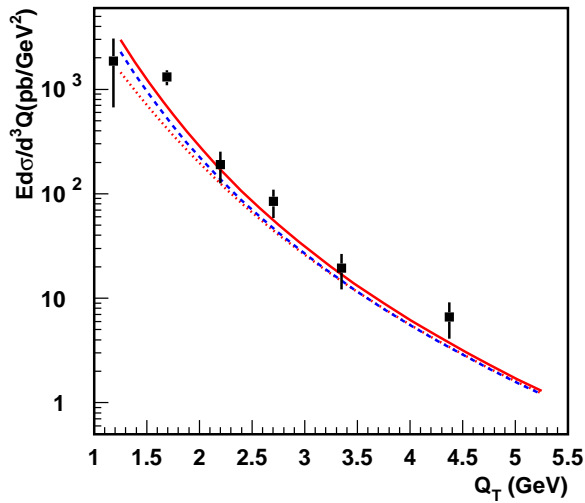, width=3.0in}
\caption{Invariant cross section for low-mass dilepton production, 
  defined in Eq.~(\ref{inv_xsec}), 
  as a function of the pair's transverse momentum
  in proton-proton collisions at $\sqrt{s}=200$~GeV
  and $y=0$, for $\kappa=1$ (solid line) and for $\kappa=0$ (dotted line).
  Also shown is the approximate cross section evaluated 
  by using Eq.~(\ref{photon}) with a NLO prompt photon
  cross section (dashed). The data are from Ref.~\cite{Akiba}.}
\label{fig5}
\end{center}
\end{figure}

In Fig.~\ref{fig5}, we show the invariant cross section 
for producing low mass electron-positron pairs, 
defined in Eq.~(\ref{inv_xsec}), 
as a function of the pair's transverse momentum, $Q_T$,
in proton-proton collisions at $\sqrt{s}=200$~GeV and
rapidity $y=0$.  We have included the NLO hard part for 
the direct contribution \cite{Berger:2001wr,BGK} and the LO
hard part for the fragmentation contribution, which 
matches our choice of a LO fragmentation function \cite{gv}. 
As mentioned earlier,
we choose $Q_{\rm min}=100$~MeV and $Q_{\rm max}=300$~MeV, 
to match the kinematics of the PHENIX measurements \cite{Akiba},
and $\lambda=200$~MeV. 
We find that setting the non-perturbative mass parameter 
$\lambda$ to zero increases the total cross section by a few 
percent.  
To test the uncertainty associated with the non-perturbative input
fragmentation functions, we plot the cross section under two 
assumptions: $\kappa=1$ (solid line) and $\kappa=0$ 
(dotted line).  The choice $\kappa=0$ corresponds to 
fragmentation functions generated without any hadronic input, 
which should be considered as the lower limit for
our numerical predictions.  On the other hand, $\kappa=1$ 
corresponds to an almost maximal hadronic 
contribution, since it amounts to neglecting all potential 
interferences between the different sources of lepton production 
and to setting the form factor to unity.  We thus expect that the 
range between the solid and the dotted line should roughly represent
the uncertainty of our prediction due to the use of our model for
the input fragmentation function, up to a few percent
additional uncertainty related to the choice of $\lambda^2$.
For comparison, we also show in Fig.~\ref{fig5}
the approximate invariant dilepton cross section 
in Eq.~(\ref{photon}), using a full NLO prompt photon cross 
section (dashed line) \cite{gv} with the parton-to-photon fragmentation
functions of~\cite{photff}. One can see that the 
latter result has the similar $Q_T$ shape as 
the one for the choice $\kappa=1$ in the full calculation,
but a smaller overall normalization.  

As we mentioned earlier, the non-perturbative parton-to-virtual-photon 
fragmentation functions have not been measured independently, and the 
normalization of the low-mass dilepton cross section calculated 
here is sensitive to the relative size of the fragmentation 
contribution.  We find that the fragmentation contribution is a fairly
important part of the production rate. It is dominated by
the QED logarithmic term in the input distribution in our 
model, see Eq.~(\ref{Dq0}).  This logarithmic $Q^2$ dependence 
is one origin of
the difference in the production rates found for virtual and real
photons in Fig.~\ref{fig5}.
The approximate invariant dilepton cross section 
in Eq.~(\ref{photon}), which is based on the full 
NLO prompt photon cross section, neglects any $Q^2$ dependence of 
the fragmentation contribution of the full calculation, which
is not very small for the range of $Q$ relevant to the PHENIX data
as indicated in Fig.~\ref{fig4}.  
The logarithmic $Q^2$ dependence of the fragmentation functions, in
combination with the factor $1/Q^2$ in Eq.~(\ref{frag_input}),
could generate a term proportional to a double logarithm
$\ln^2(Q^2_{\rm max}/Q^2_{\rm min})$ and enhance the overall 
production rate in comparison to the single logarithm 
$\ln(Q^2_{\rm max}/Q^2_{\rm min})$ in Eq.~(\ref{photon}). 
Also, our full calculation uses only the LO hard part
and LO parton-to-virtual-photon fragmentation functions.  
Using a NLO hard part and NLO parton-to-virtual-photon 
fragmentation functions, which are not available yet, 
would be expected to further increase our predicted production 
rate.  In any case, the difference between the solid line
and the dashed (real-photon) line in Fig.~\ref{fig5} should be roughly
representative of the present theoretical uncertainty of the calculation 
of the low-mass dilepton cross section for the given kinematics.

\bef
\psfig{file=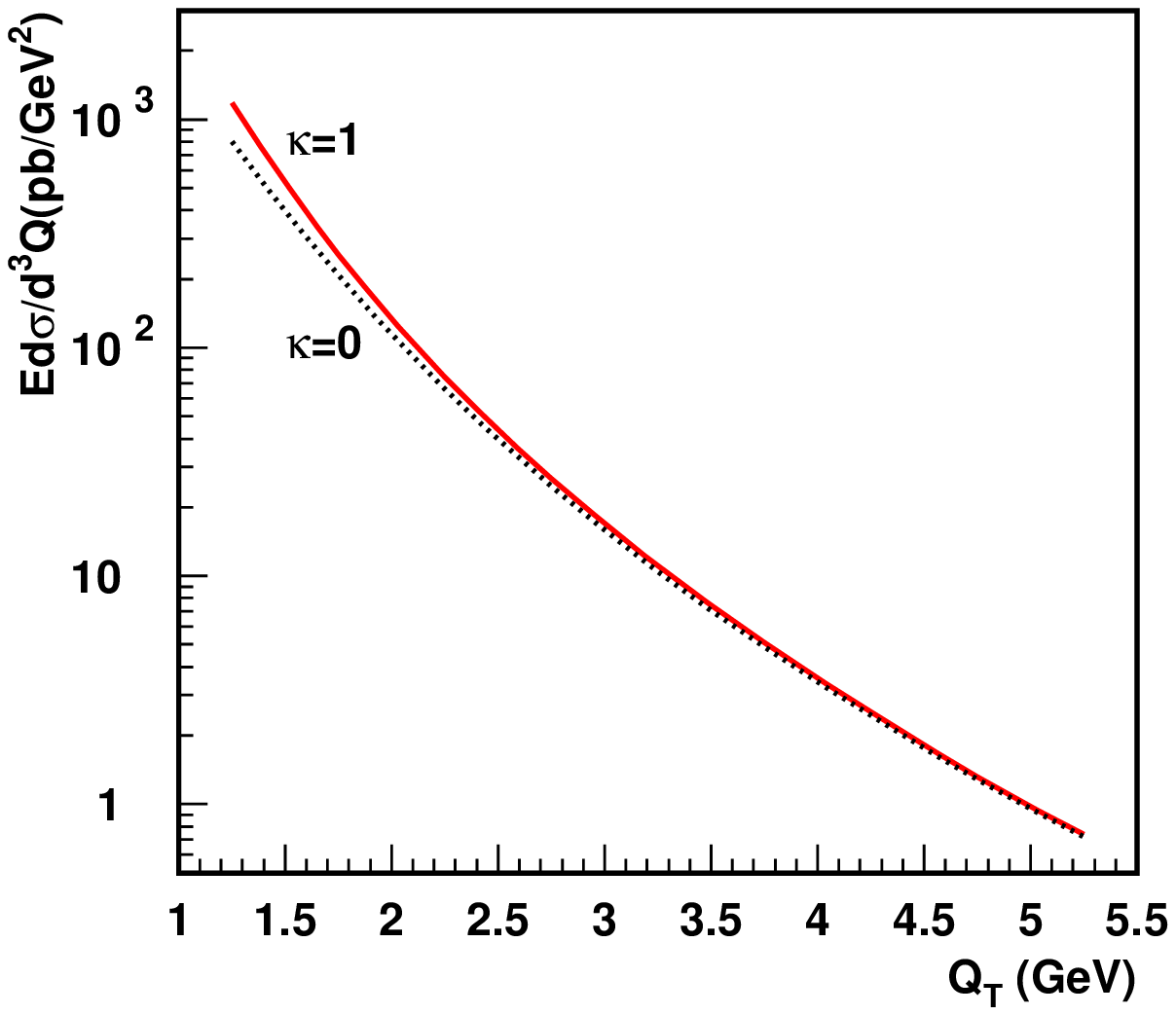,width=3.0in}
\hskip 0.2in
\psfig{file=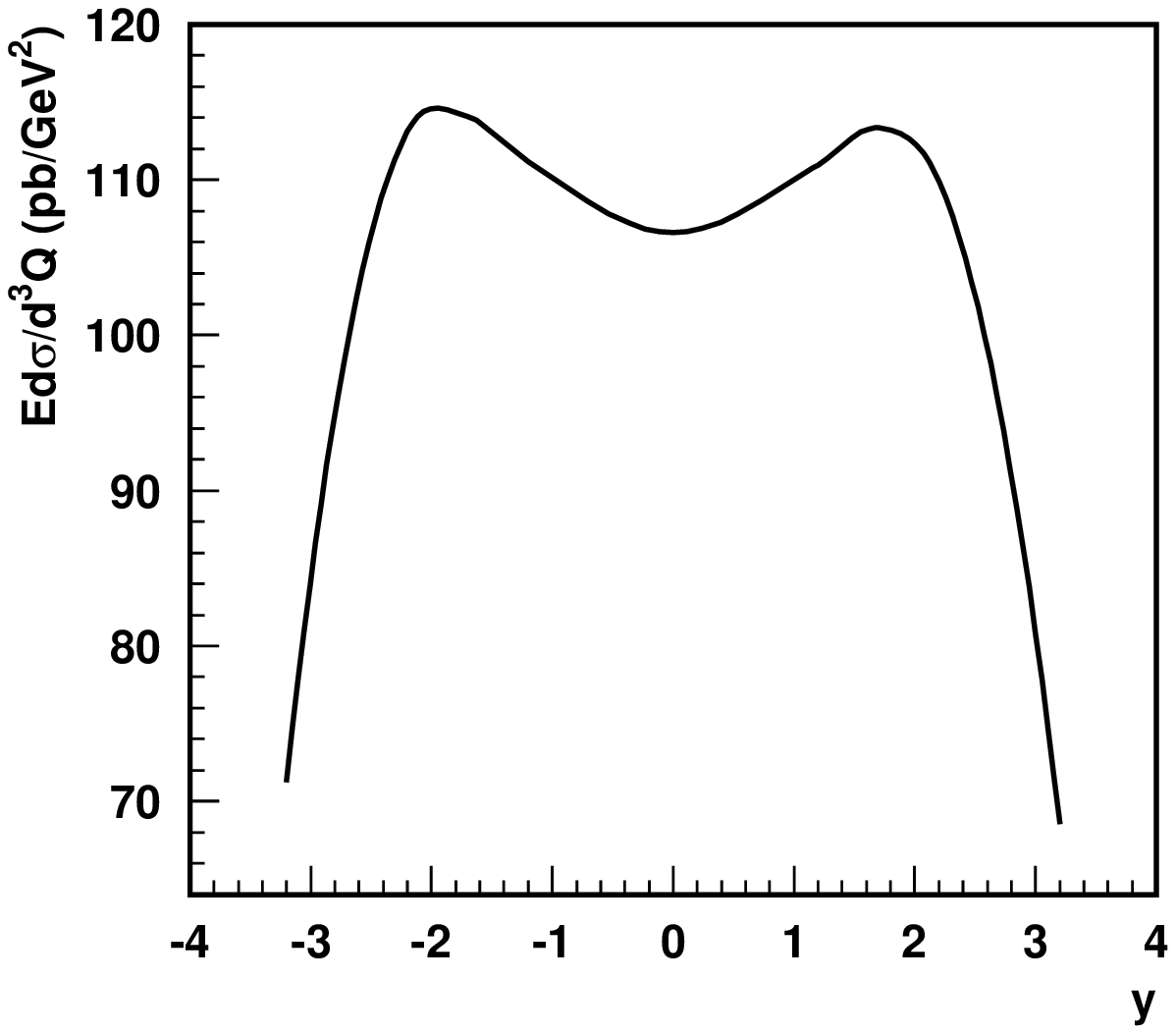,width=3.0in}
\caption{Invariant cross section for low mass $\mu^+\mu^-$ production 
as a function of the transverse momentum $Q_T$ at $y=1.7$, and 
as a function of the rapidity at $Q_T=2$ GeV.
The solid line is for $\kappa=1$ and the dotted line is for $\kappa=0$.}
\label{dimuon}
\eef

In Fig.~\ref{dimuon}, we plot the invariant cross section 
for low-mass dimuon production in proton-proton collisions 
at $\sqrt{s}=200$~GeV. We choose here the range $Q_{\rm min}=$250~MeV 
to $Q_{\rm max}=550$~MeV, so that we avoid both the mass threshold
at $Q=2 m_\mu\approx 211$~MeV and the $\rho$-meson resonance.
We plot the invariant cross section as a function of 
the dimuon's transverse momentum $Q_T$ at a forward rapidity 
$y=1.7$, which is roughly the average rapidity covered by the 
PHENIX muon arm. We again present two curves: the solid line 
corresponds to $\kappa=1$ and the dotted one to $\kappa=0$. In the
right part of the figure, we show the invariant cross section as a
function of $y$ at $Q_T=2$~GeV and $\kappa=0$.  
Because of the larger invariant mass and more forward angles 
we are considering here,
the production rate is smaller than in the electron-positron case.

\subsection{Isospin effects in nuclear low-mass dilepton production}

Like prompt photon production, the hadronic production
of low-mass lepton pairs is also dominated by the 
$qg\to \gamma^* q$ Compton subprocess.  Therefore, the invariant
cross section for low-mass dilepton production could have
significant isospin effects in nuclear collisions, similar to the
ones observed in recent studies of the real photon cross 
section \cite{arleo,iso_gamma}.

In order to present nuclear effects in high energy nuclear 
collisions, one often defines a nuclear modification factor
as the ratio of the nuclear cross section, 
normalized to the number of binary collisions, over the proton-proton 
one. As the proton-proton cross section differs from that
for proton-neutron or neutron-neutron scattering,
the nuclear modification factor will be different from unity
simply by isospin effects, even in the absence of any ``genuine'' 
nuclear effects. To quantify the size of the isospin effect in 
deuteron-gold collisions, we introduce the ratio 
\ben
R_{\mathrm{d Au}}^{\mathrm{iso}}\equiv
\frac{\frac{1}{2A}d^2\sigma^{\mathrm{d Au}}/dQ_Tdy}{d^2\sigma^{pp}/dQ_Tdy}\, ,
\label{RdA}
\een
where the deuteron-gold cross section in the numerator is calculated
by replacing the proton's parton distribution functions (PDFs) by
\ben
f_i^{p}(x, Q^2)\rightarrow 
\left[Z\cdot f_i^{p}+(A-Z)\cdot f_i^{n}
\right]/A\, ,
\label{isospin}
\een
in the $pp$ calculation, 
with $A=197$ ($A=2$) and $Z=79$ ($Z=1$) the atomic weight and proton number
of the gold (deuteron) nucleus, respectively. For gold-gold 
collisions, we introduce a corresponding 
$R^{\mathrm{iso}}_{\mathrm{Au Au}}$.

\bef
\psfig{file=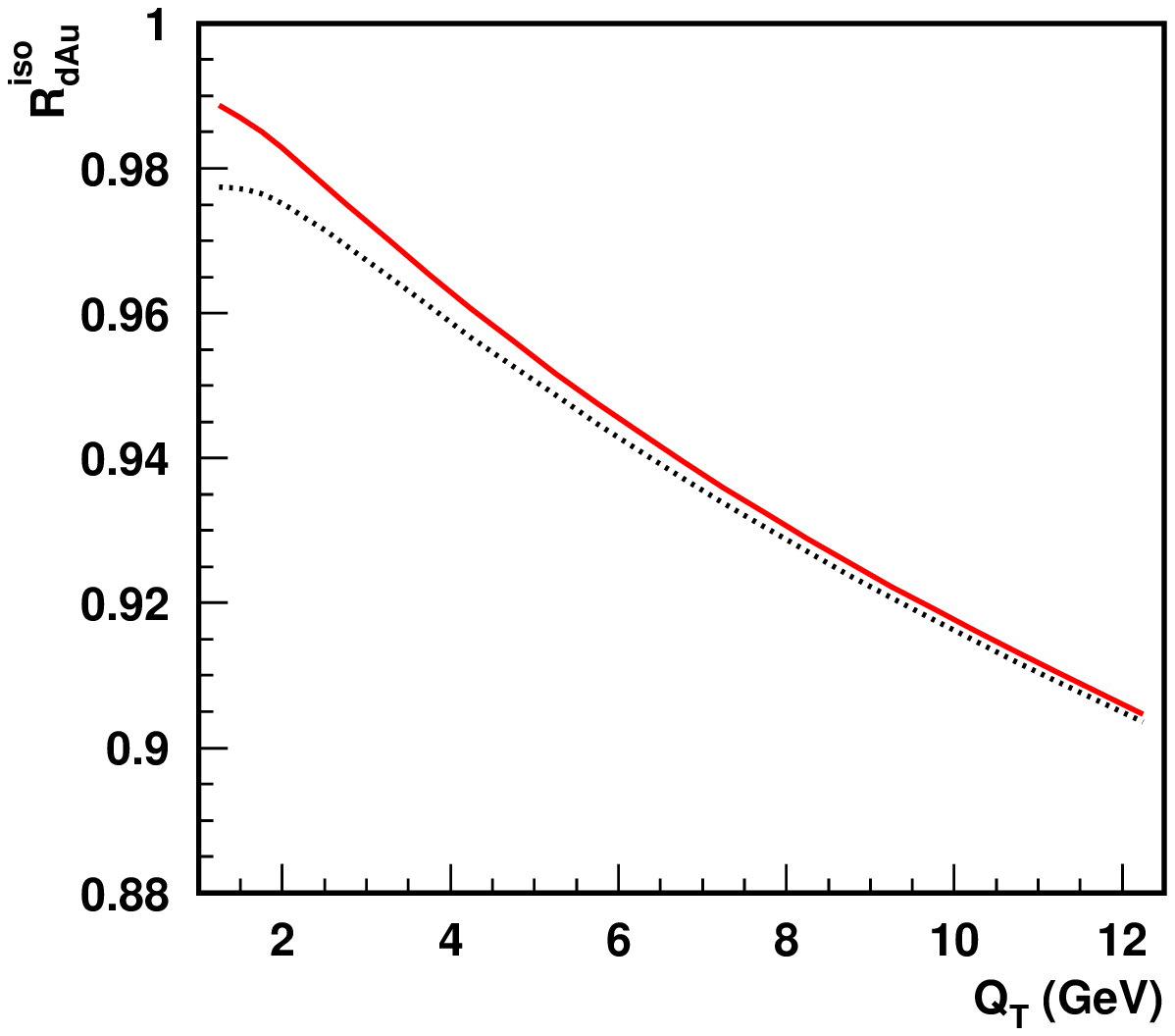,width=3.0in}
\hskip 0.2in
\psfig{file=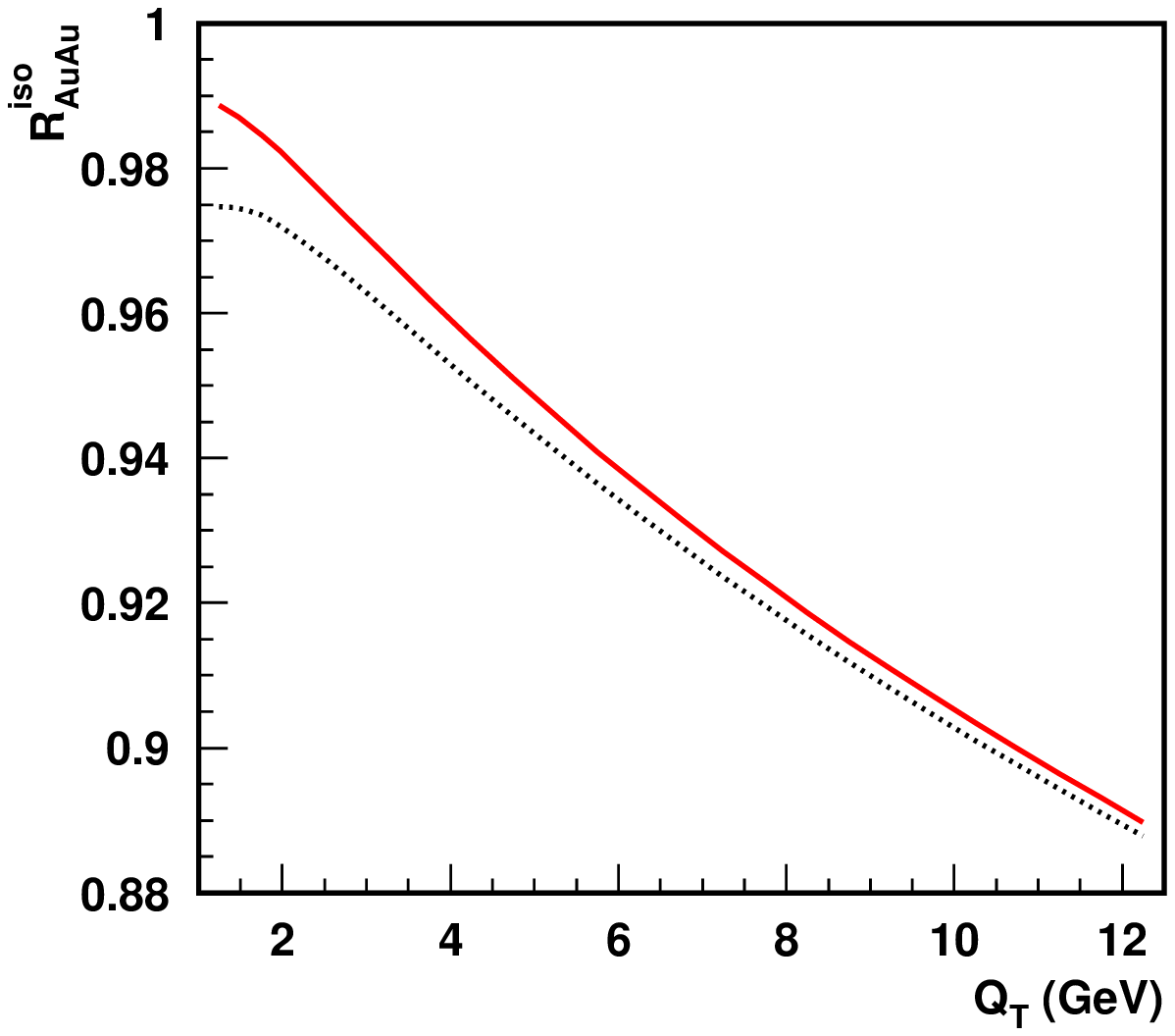,width=3.0in}
\caption{$R^{\mathrm{iso}}_{\mathrm{d Au}}$ and 
$R^{\mathrm{iso}}_{\mathrm{Au Au}}$ as functions of $Q_T$
 at $\sqrt{s}=200$~GeV and $y=0$: $\kappa=1$ (solid), 
 $\kappa=0$ (dotted).}
\label{isospin_more}
\eef

In Fig.~\ref{isospin_more}, we plot $R_{\mathrm{d Au}}^{\mathrm{iso}}$ and 
$R_{\mathrm{Au Au}}^{\mathrm{iso}}$ for low-mass electron-positron pair
production as functions of $Q_T$. As before, we integrate over the
mass range $100 \leq Q \leq 300$~MeV. The solid lines are for 
$\kappa=1$ while the dotted ones are for $\kappa=0$.
We find that the isospin effect is significant, giving rise to a reduction of
the nuclear modification factor by as much as 10\% at $Q_T\sim 10$~GeV.  
This finding is consistent with the results obtained for real photon
production in~\cite{arleo,iso_gamma}.
This reduction is readily understood: for low-mass dilepton production
at large transverse momentum, the $qg\to \gamma^*q$ 
Compton subprocess dominates the production rate.  
In the case of proton-proton collisions, the Compton process
involves the quark combination $\frac{4}{9}f_u^p+\frac{1}{9}f_d^p$.
On the other hand, for the case of neutron-neutron collisions, 
this combination becomes
$\frac{4}{9}f_u^n+\frac{1}{9}f_d^n=\frac{4}{9}f_d^p+\frac{1}{9}f_u^p$. 
Since $f_u^p>f_d^p$, the invariant cross section at central rapidities
follows the relation: $\sigma^{nn}<\sigma^{np}=\sigma^{pn}<\sigma^{pp}$.
As a result, $R_{AB}^{\mathrm{iso}}<1$ for any nucleus-nucleus collision.
In addition, the stronger suppression with increasing $Q_T$ is
caused by the fact that the ratio $f_d^p(x)/f_u^p(x)$ 
decreases as $x$ increases.

The dependence of the results on $\kappa$ in Fig.~\ref{isospin_more}
arises in the following way. When $\kappa=1$, the fragmentation 
contribution is much larger.  Since this contribution has a 
substantial contribution from the gluon-gluon fusion subprocess, it 
is much less sensitive to any isospin effect.  
Consequently, for the $\kappa=1$ case one finds a smaller 
isospin suppression at low $Q_T$, whereas at higher $Q_T$,
where the fragmentation contribution becomes overall less important, 
the results become independent of $\kappa$.

\subsection{Low-mass lepton pairs in nuclear collisions}
\label{nuclear}

Since the cross section for low-mass lepton pairs 
at large transverse momentum 
is dominated by the $qg$ Compton subprocesses, 
it is expected to be an excellent probe of the gluon distribution 
function \cite{Berger:2001wr,Berger:1998ev}, 
and its nuclear dependence.
In this subsection, we investigate this issue by studying the dependence
of the cross section on the choice of nuclear parton distribution functions 
(nPDFs). We neglect any other nuclear effects, such as saturation effects
\cite{saturation},
parton multiple scattering in the nuclear medium 
\cite{Guo:1997it,parton_m_col,iso_gamma}, 
or thermal radiation of lepton pairs~\cite{thermal}.  
The nPDFs, or more precisely, 
the effective PDFs of a bound proton inside a nucleus of 
atomic weight $A$ are often written as  
\ben
f_{i}^{p/A}(x,Q^2) \equiv \rho_{i}^A(x,Q^2) f_{i}^{p}(x,Q^2)\, 
\label{nPDF}
\een
where the factors $\rho_i^A(x,Q^2)$ are usually determined
in global QCD fits to the data from deep inelastic lepton
scattering off nuclei \cite{EKS98,fgs,kumano,dFS2003,EPS08}. The  
nuclear PDFs for a bound neutron are obtained 
from those for the proton by using isospin symmetry. For example,
$f_u^{n/A} = f_d^{p/A}=\rho_d^A(x,Q^2) f_d^{p}$. Following 
Eq.~(\ref{isospin}) we then obtain
the parton distribution for flavor $i$ in the nucleus as
\ben
\left[Z\cdot f_i^{p/A}+(A-Z)\cdot f_i^{n/A}
\right]/A\, .
\label{npdf}
\een

The nuclear modification factor for deuteron-nucleus collisions 
is defined as
\ben
R_{\mathrm{d Au}}\equiv
\frac{1}{\langle N_{coll}\rangle}\frac{d^2N^{\mathrm{d Au}}/dQ_Tdy}
{d^2N^{pp}/dQ_Tdy}
 \stackrel{\rm min. bias}{=}
 \frac{\frac{1}{2A}d^2\sigma^{\mathrm{d Au}}/dQ_Tdy}{d^2\sigma^{pp}/dQ_Tdy}\, ,
\een
where $\langle N_{coll}\rangle$ is the number of binary collisions and
the label ``min. bias'' refers to minimum bias events.
The modification factor for gold-gold collisions, $R_{\mathrm{Au Au}}$, is
defined analogously. 

\bef
\psfig{file=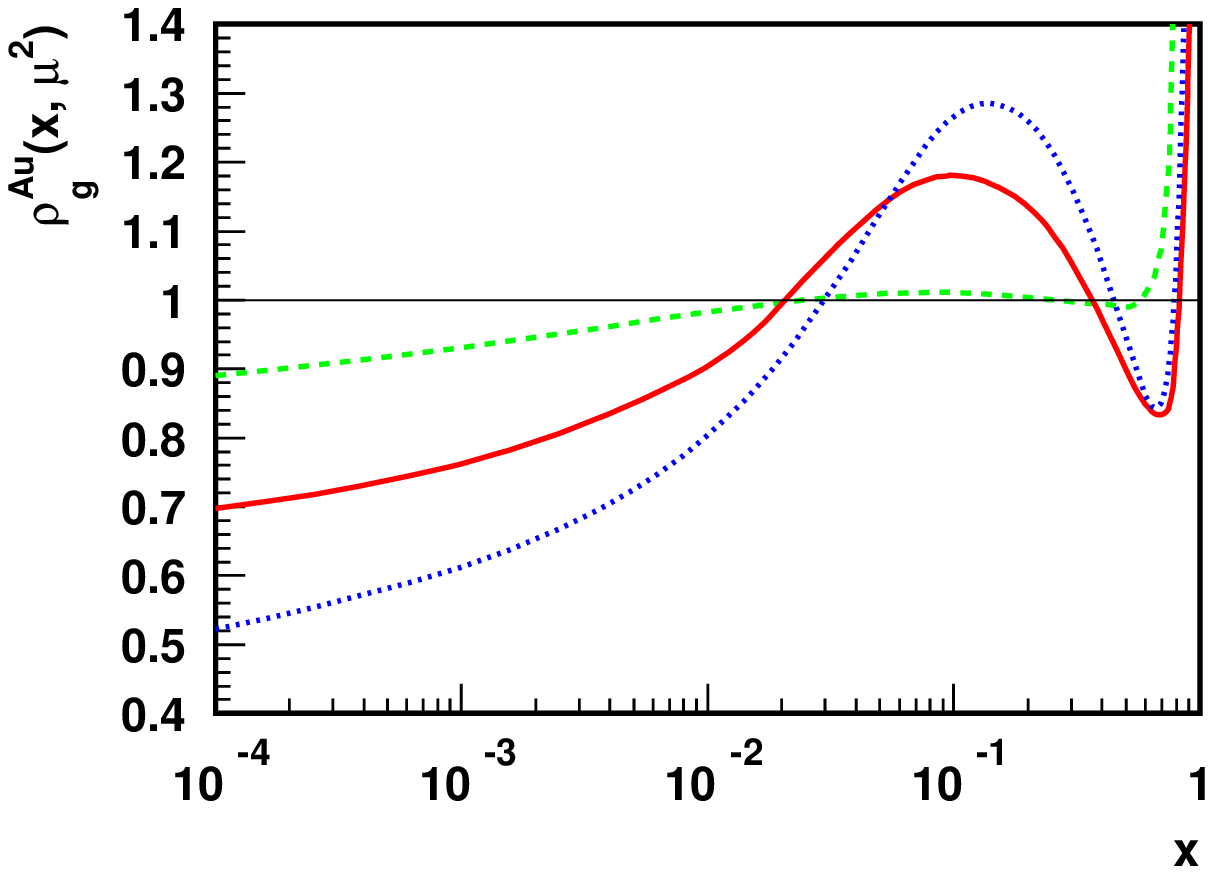, width=3.0in}
\caption{The ratio $\rho_g^{\mathrm{Au}}(x,\mu^2)$
of the gluon distributions in a gold nucleus and 
in the proton, defined in Eq.~(\ref{nPDF}), as a function of $x$
at $\mu=3$~GeV, evaluated for three sets of nPDFs: 
EKS98 (solid), EPS08 (dotted), and dFS2003 (dashed).}
\label{Rgluon}
\eef

In order to investigate the sensitivity of the nuclear modification factors
to the nuclear gluon distribution, we use three
sets of nPDFs in our plots below: EKS98 \cite{EKS98}, 
dFS2003 \cite{dFS2003}, and the recent EPS08 set \cite{EPS08}.  
We emphasize that none of the three sets considers the dependence
of the nPDFs on the transverse location inside the nucleus
(the impact parameter dependence) \cite{impact}.
Using Eq.~(\ref{nPDF}) implies that we average over
the impact parameter dependence of the nuclear effects.
While all sets fit the data from which the nPDFs were extracted, 
they predict a significantly different gluon distribution function
in a large nucleus, as shown in Fig.~\ref{Rgluon}.
While the gluon distribution of dFS2003 has an overall relatively weak 
nuclear dependence, the other two sets have strong shadowing 
at small $x$ and anti-shadowing at $x\sim 0.1$, in particular
for EPS08.

\bef
\psfig{file=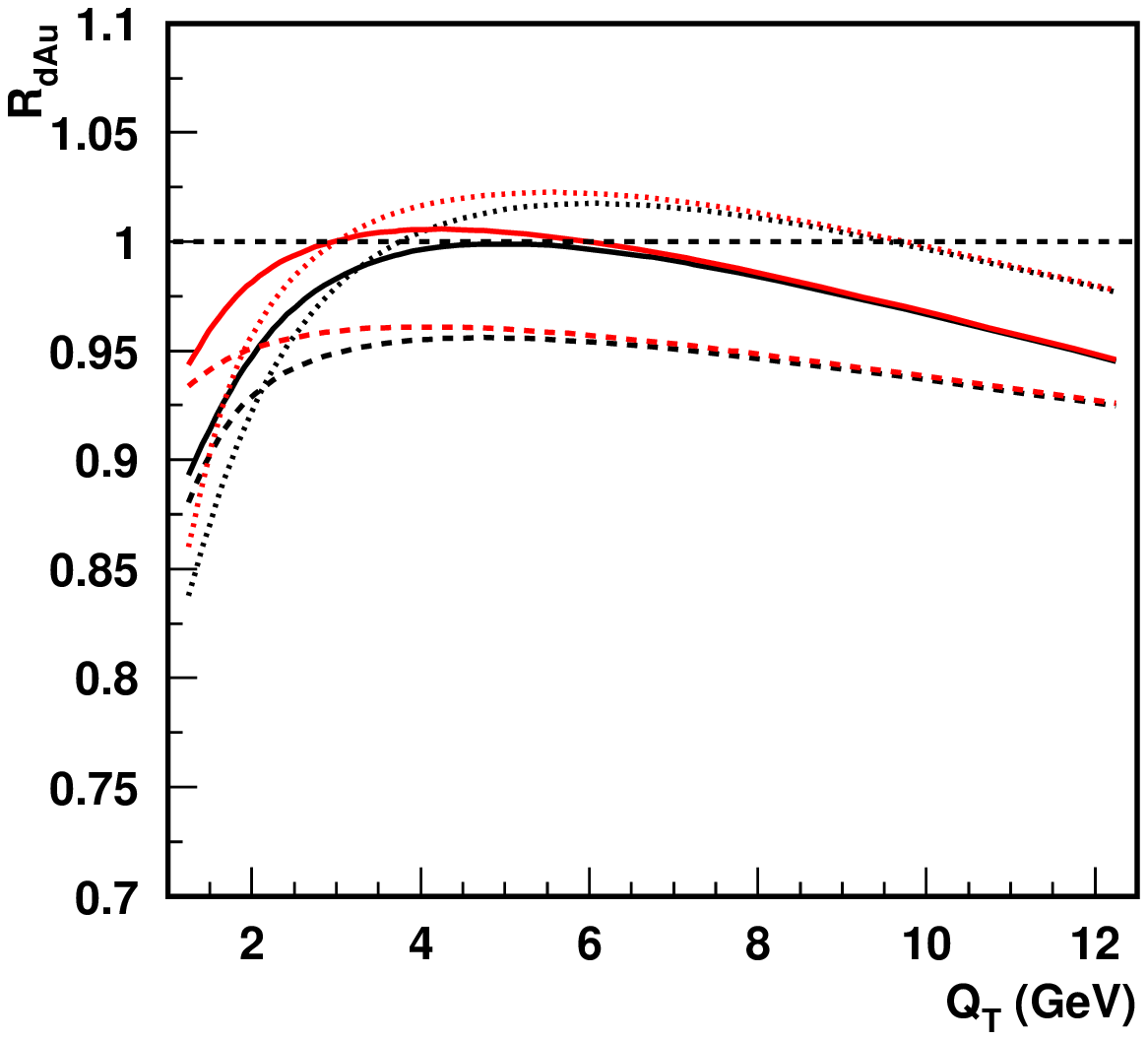,width=3.0in}
\hskip 0.2in
\psfig{file=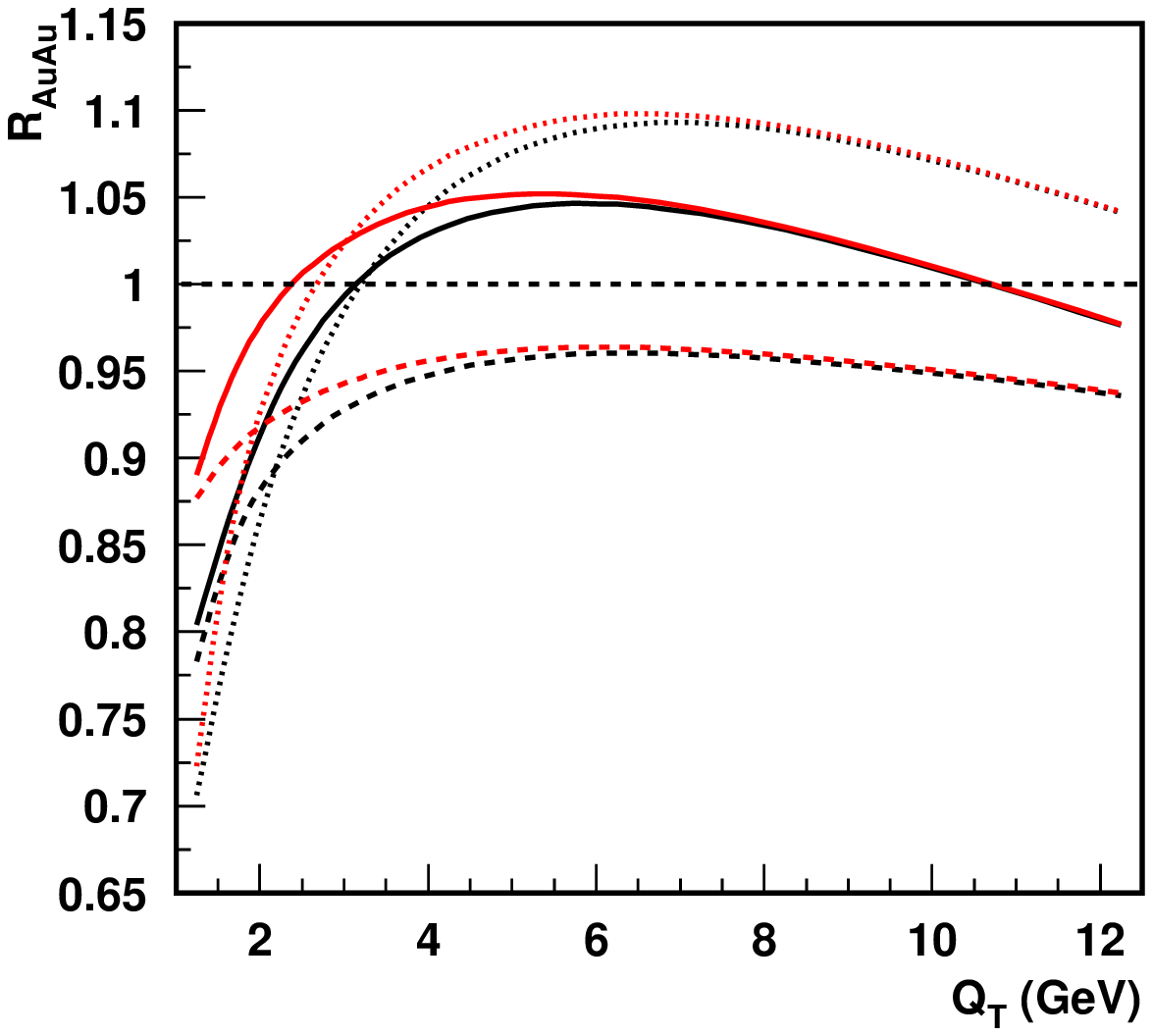,width=3.0in}
\caption{The nuclear modification factor for the invariant 
cross section for low-mass dilepton production, 
$R_{\mathrm{d Au}}$ and $R_{\mathrm{Au Au}}$, as functions of $Q_T$ 
at $\sqrt{s}=200$~GeV and $y=0$.  
The solid, dashed, and dotted lines correspond
to the EKS98, dFS2003, and EPS08 sets of nPDFs, respectively.  
For each set, the top curve (red color online) is for $\kappa=1$ and 
the bottom one (black color online) for $\kappa=0$.}
\label{nuclear_more}
\eef

\bef
\psfig{file=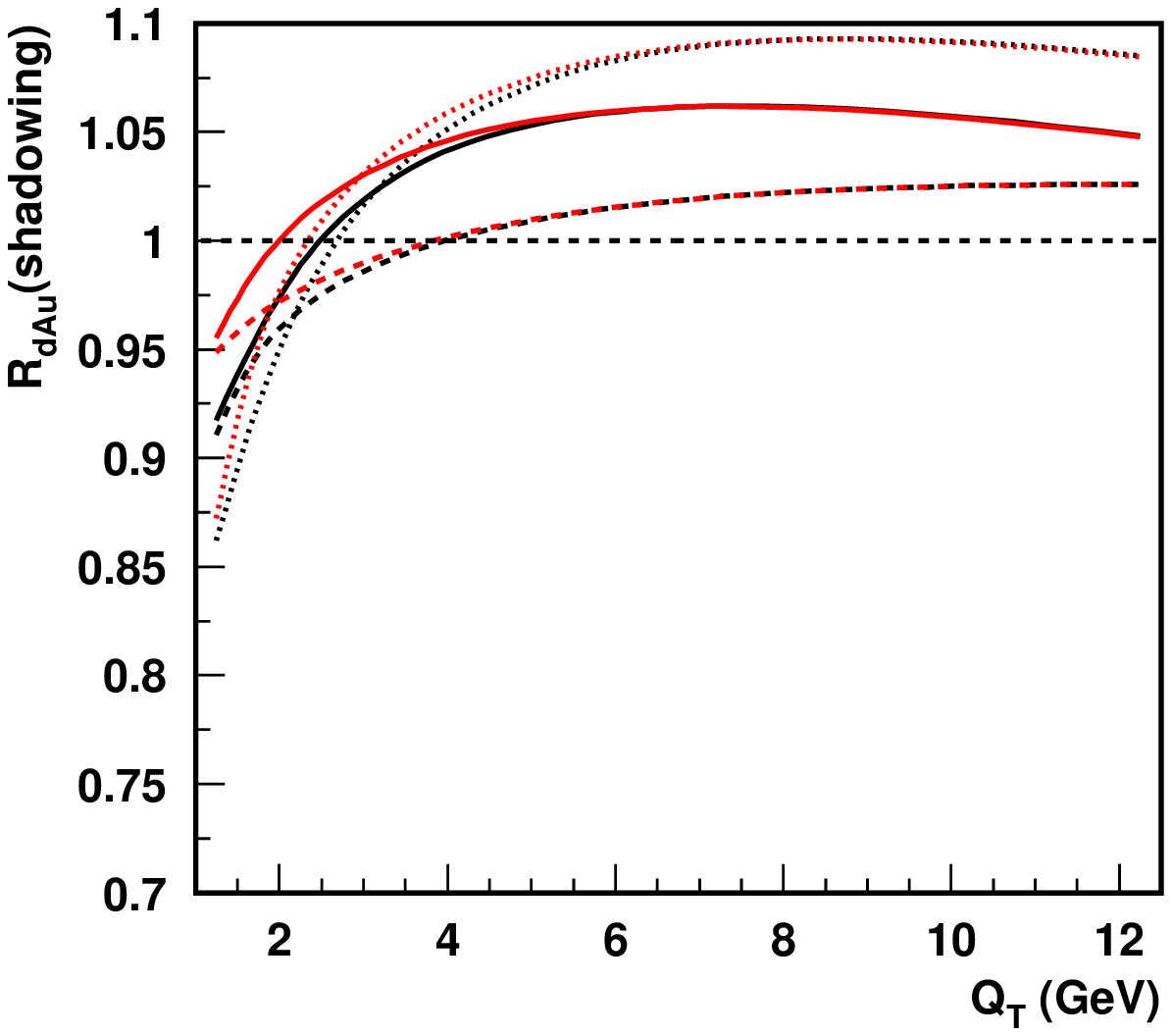,width=3.0in}
\hskip 0.2in
\psfig{file=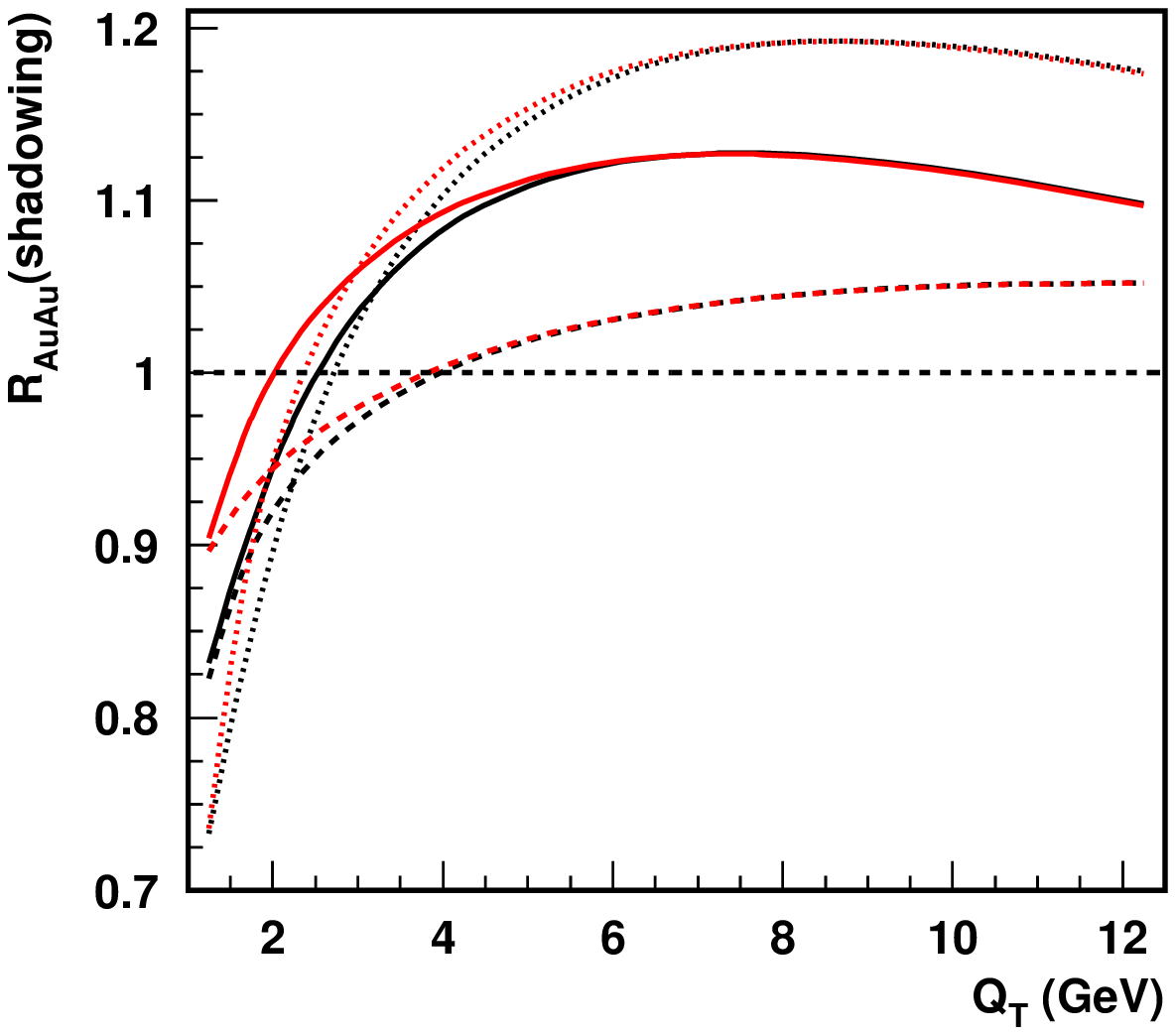,width=3.0in}
\caption{Same as in Fig.~\ref{nuclear_more}, but ignoring the
isospin effects of the nuclear PDFs.
These plots thus represent the nuclear shadowing and anti-shadowing
effects in isolation.}
\label{pureshadow}
\eef

In Fig.~\ref{nuclear_more}, we plot the nuclear modification factors
$R_{\mathrm{d Au}}$ and $R_{\mathrm{Au Au}}$ for the invariant 
cross section for low-mass electron-positron pair production in 
nuclear collisions as functions of the lepton pair's transverse momentum, 
at $\sqrt{s}=200$~GeV and $y=0$. As expected from the differences
in the nuclear gluon distribution shown in Fig.~\ref{Rgluon},
the three sets of nPDFs lead to significant differences in the 
predictions for the nuclear modification factors. One can
first of all see that the EPS08 set gives modification
factors that become larger than unity at $Q_T\gtrsim 3$~GeV. This implies 
that for the kinematics we are considering the anti-shadowing region 
dominates, which as we saw is most pronounced for the EPS08 set.
The EKS98 set with its slightly lower anti-shadowing still leads to 
$R_{\mathrm{Au Au}}>1$ at $Q_T\gtrsim 3$~GeV, while for dFS2003, 
which hardly has any anti-shadowing of the gluon distribution at
all, both $R_{\mathrm{d Au}}$ and $R_{\mathrm{Au Au}}$ remain below 
unity. Since at mid-rapidity the momentum fractions of the two
colliding partons are on average the same, it is not surprising that
the anti-shadowing contributions are amplified in $R_{\mathrm{Au Au}}$,
as shown by $R_{\mathrm{Au Au}}>R_{\mathrm{d Au}}$ at high $Q_T$.
We thus conclude that the nuclear modification factor for the production of
low-mass electron-positron pairs at RHIC may provide 
valuable information on the nuclear gluon distribution. We note that 
of course the nuclear effects in the quark (and anti-quark) distributions
also matter for the precise behavior of $R_{\mathrm{d Au}}$ and 
$R_{\mathrm{Au Au}}$; however, these distributions are much better
known than the gluon one and are more similar in the various nPDF sets.

For all three sets of nPDFs we consider, $R_{\mathrm{d Au}}$ and 
$R_{\mathrm{Au Au}}$ decrease at very high $Q_T$ and tend to 
go below unity. The nuclear modification factors evaluated using 
the dFS2003 set actually remain below unity in the entire range of $Q_T$.  
This turns out to be a result of the isospin effect we discussed
earlier, which pushes the nuclear modification factors to lower
values. In order to verify this, we plot in Fig.~\ref{pureshadow}
the same nuclear modification factors, evaluated however ignoring 
the isospin effect by setting $A=Z$ in Eq.~(\ref{npdf}). As expected,
the nuclear modification factor in Fig.~\ref{pureshadow} then
follows the features of the nuclear gluon distribution more closely.

\bef
\psfig{file=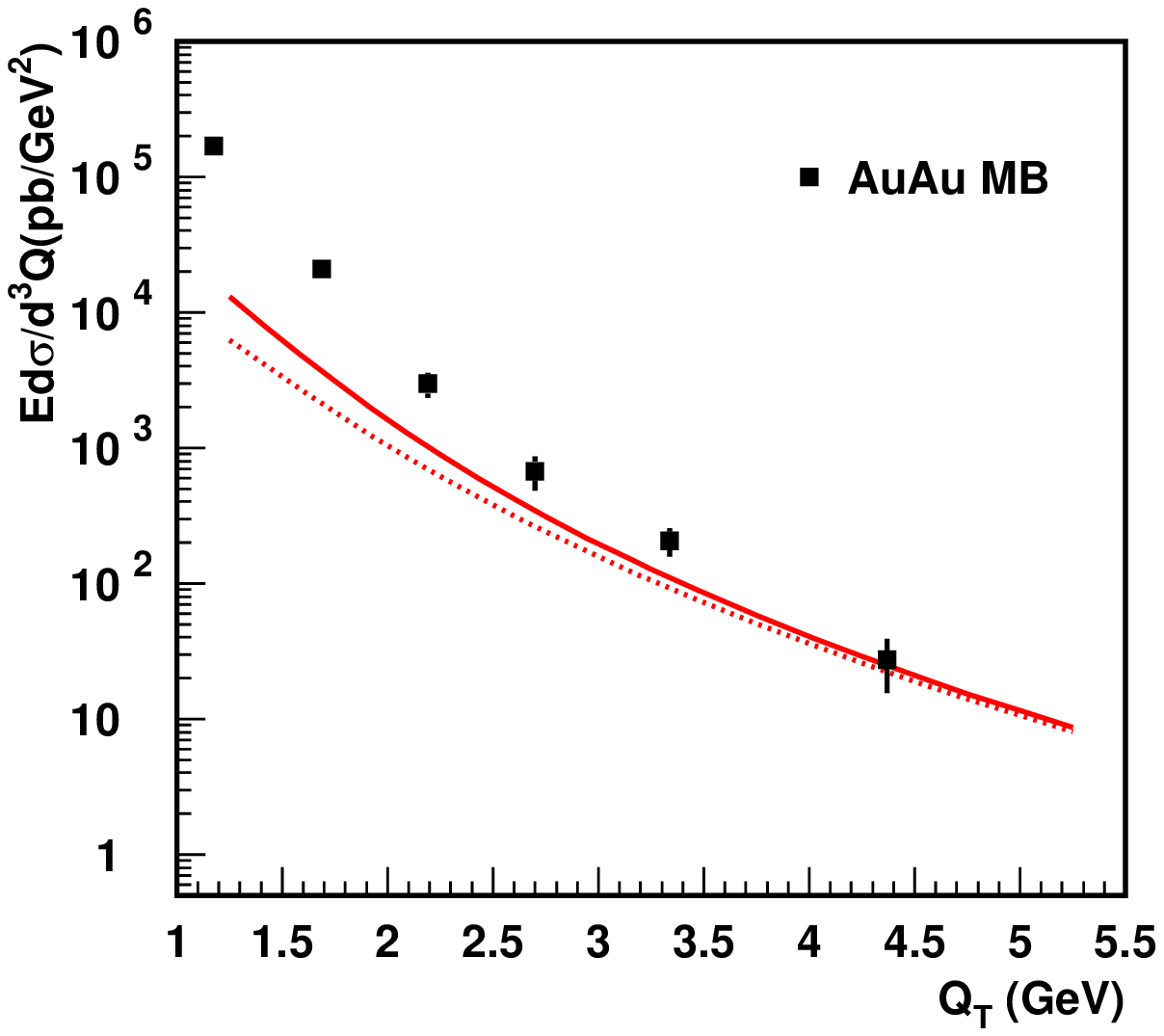,width=3.0in}
\psfig{file=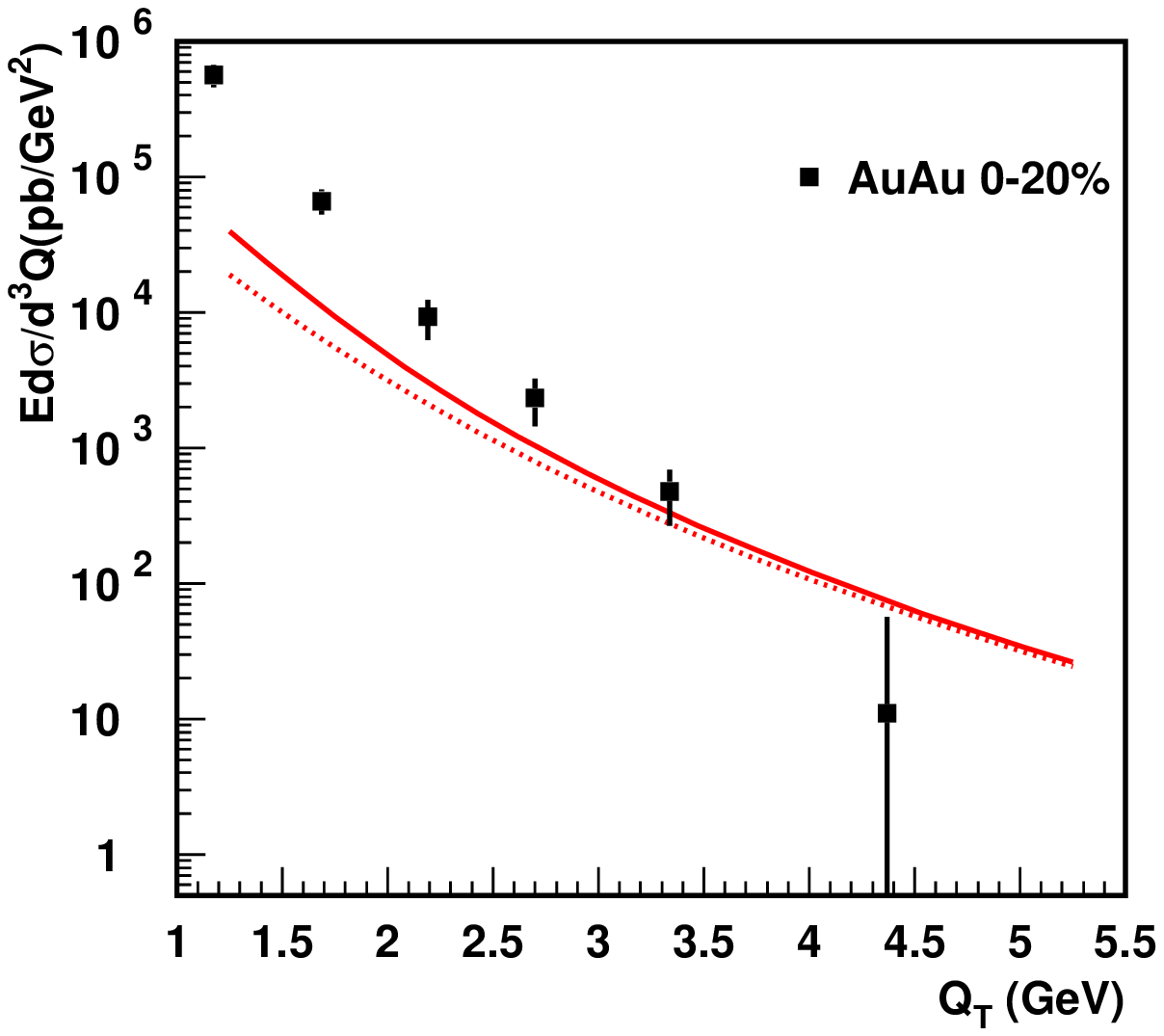,width=3.0in}
\psfig{file=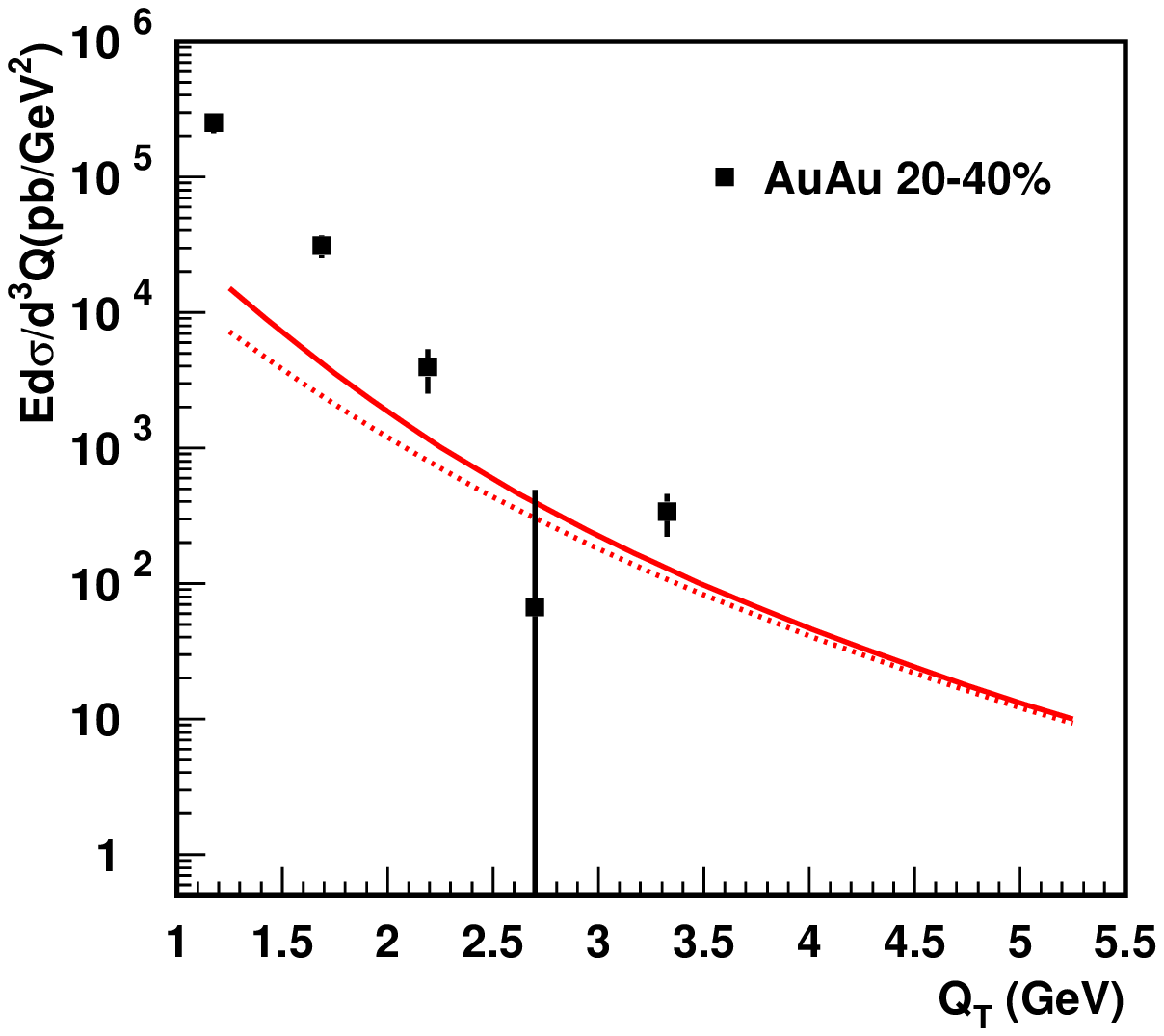,width=3.0in}
\caption{Invariant cross section for low-mass dilepton production
in Au+Au collisions as a function of the pair's transverse momentum,
evaluated by using Eq.~(\ref{inv_xsec}) and the EPS08 nPDFs 
at $\sqrt{s}=200$~GeV and $y=0$. We use $\kappa=1$ (solid) 
and $\kappa=0$ (dotted).
The centrality dependence of nuclear collisions is taken
into account by scaling the cross section with 
the Glauber nuclear overlap function \cite{TAA}.  
Experimental data are from Ref.~\cite{Akiba}.}
\label{fig11}
\eef


We note that some of the features we find have also been 
observed in studies of the real photon cross section \cite{arleo,iso_gamma}.
We remind the reader that the importance of the low-mass dilepton
observable lies in the fact that it allows to extend the measurements
to lower $Q_T\lesssim 5$~GeV. For real prompt photons, preliminary 
PHENIX measurements find $R_{\mathrm{d Au}}$~\cite{Peressounko:2006qs} 
and $R_{\mathrm{Au Au}}$~\cite{KleinBosing:2007bp} to be consistent with 
unity within relatively large errors, with $R_{\mathrm{Au Au}}$ appearing 
to drop to lower values at $Q_T\gtrsim 12$~GeV.

Finally, we close this section by showing in Fig.~\ref{fig11} a comparison 
of our calculated low-mass dilepton cross section 
in Au+Au collisions with the recent RHIC data at $\sqrt{s}=200$~GeV and 
$y=0$ \cite{Akiba}. We use Eq.~(\ref{inv_xsec}), the EPS08 nPDFs, and 
$\kappa=1$ (solid line), $\kappa=0$ (dotted line).
We take into account the centrality dependence of nuclear collisions 
by scaling the calculated cross section with the 
average Glauber nuclear overlap function
$\langle T_{AA} \rangle$, which is the same as that used by
the PHENIX experiments \cite{TAA}.
We find that thanks to the anti-shadowing effect
the calculated cross section is roughly consistent with the
experimental data for $Q_T>3$~GeV.  Without including 
other potential nuclear effects, such as power corrections from 
parton multiple scattering \cite{Guo:1997it}
or thermal radiation~\cite{thermal}, which could enhance the dilepton 
cross section at low $Q_T$, our calculated cross section 
in Au+Au collisions with the nuclear modification from nPDFs alone 
cannot explain the existing RHIC data.  
As shown by Fig.~\ref{fig11}, the production rate 
of low-mass lepton pairs in Au+Au collisions 
measured by PHENIX is clearly larger than 
the perturbative QCD calculation for $Q_T\lesssim 3$~GeV. 
As pointed out in Ref.~\cite{arleo}, energy loss of partons
in the fragmentation contribution could generate additional nuclear
dependence of the direct photon cross section, which could reduce the production
rate in Au+Au collisions. We expect the effect of energy loss of the
fragmentating partons on low mass dilepton production to be similar.

\section{Summary}
\label{summary}

We have investigated the production cross section 
for lepton pairs in the regime $Q_T \gg Q \sim \Lambda_{\rm QCD}$.
We have argued that, like for real photons, the cross section 
can be systematically factorized into universal fragmentation
functions, parton distributions,
and perturbatively calculable partonic hard parts 
evaluated at a distance scale $\sim {\cal O}(1/Q_T)$. 
We conclude that the lepton-pair cross section in this 
kinematic regime can be treated in a similar fashion, 
and with similar rigor, as the real-photon one.

When $Q \sim \Lambda_{\rm QCD}$, the fragmentation contribution
to the lepton-pair cross section cannot be completely perturbative 
as it is for $Q \gg \Lambda_{\rm QCD}$. 
We have introduced models for the lepton pair input fragmentation 
functions at a scale $\mu_0\sim 1$~GeV, which we have evolved 
perturbatively to higher scales. Using the evolved fragmentation functions,
we have calculated the transverse momentum distributions of low-mass 
lepton pairs in proton-proton collisions at $\sqrt{s}=200$~GeV at RHIC.
Our calculated cross section is consistent with existing RHIC data
\cite{Akiba}.  A careful comparison with the existing and future data 
should help to better pin down the universal fragmentation functions.  

We have also discussed the case of nuclear collisions at RHIC, and 
have found that the nuclear modification factors in deuteron-gold
and gold-gold collisions for high-$Q_T$ low-mass dilepton production
are good probes of the nuclear gluon distribution, similar to those
for real photon production. We also noticed that isospin effects are 
very significant for the nuclear modification factors. 
We demonstrated that our calculated cross section in Au+Au collisions 
with the nuclear modification from nPDFs alone cannot explain 
the existing RHIC data.  The data from Au+Au collisions show a clear 
enhancement over the perturbative QCD calculation for $Q_T\lesssim 3$~GeV.

We finally stress that the production of low-mass lepton pairs and 
the prompt photons at large transverse momentum in 
high energy hadronic collisions have many features in common. 
However, experimentally, they are also complementary to each 
other, since they can be exploited in distinct kinematic regimes.  
A combination of these two electromagnetic probes 
of hard scattering could provide good information on 
gluon distributions and short-distance QCD dynamics.

\section*{Acknowledgments}

We thank Y. Akiba for many useful discussions on hadronic production
of low-mass lepton pairs, and for his careful reading and valuable comments 
on our manuscript. We are grateful to E.L. Berger for discussions
on Drell-Yan production of low-mass lepton pairs and to M. Strikman and
R. Venugopalan for discussions on nuclear parton distributions. 
This work was supported in part by the 
U.S. Department of Energy under grant number DE-FG02-87ER40371 (JQ)
and contract number DE-AC02-98CH10886 (WV).
JQ thanks the Institute of High Energy Physics, Chinese Academy
of Science for its hospitality during the writing of this work.



\begin{thebibliography}{99}

\bibitem{prompt:photon}
for reviews, see: J.~F.~Owens,
  Rev.\ Mod.\ Phys.\  {\bf 59}, 465 (1987);
W.~Vogelsang and M.~R.~Whalley,
  J.\ Phys.\ G {\bf 23}, A1 (1997).

\bibitem{Aurenche:1988yr}
  P.~Aurenche, R.~Baier and M.~Fontannaz,
  Phys.\ Lett.\  B {\bf 209}, 375 (1988).

\bibitem{Berger:1998ev}
  E.~L.~Berger, L.~E.~Gordon and M.~Klasen,
  Phys.\ Rev.\  D {\bf 58}, 074012 (1998)
  [arXiv:hep-ph/9803387].
  
\bibitem{Qiu:2001nr}
  J.~W.~Qiu and X.~f.~Zhang,
  Phys.\ Rev.\  D {\bf 64}, 074007 (2001)
  [arXiv:hep-ph/0101004].

\bibitem{Berger:2001wr}
  E.~L.~Berger, J.~W.~Qiu and X.~f.~Zhang,
  Phys.\ Rev.\  D {\bf 65}, 034006 (2002)
  [arXiv:hep-ph/0107309].

\bibitem{photff} 
M.~Gl\"{u}ck, E.~Reya and A.~Vogt,
  Phys.\ Rev.\  D {\bf 48}, 116 (1993)
  [Erratum-ibid.\  D {\bf 51}, 1427 (1995)].

\bibitem{photff1} 
L.~Bourhis, M.~Fontannaz and J.~P.~Guillet,
  Eur.\ Phys.\ J.\  C {\bf 2}, 529 (1998)
  [arXiv:hep-ph/9704447].

\bibitem{photff2} 
A.~Gehrmann-De Ridder and E.~W.~N.~Glover,
  Nucl.\ Phys.\  B {\bf 517}, 269 (1998)
  [arXiv:hep-ph/9707224].

\bibitem{Akiba} 
  A.~Adare {\it et al.}  [PHENIX Collaboration],
  arXiv:0804.4168 [nucl-ex].

\bibitem{virtphot} T.~Uematsu and T.~F.~Walsh,
  Phys.\ Lett.\  B {\bf 101}, 263 (1981); 
Nucl.\ Phys.\  B {\bf 199}, 93 (1982);
W.~Ibes and T.~F.~Walsh,
  Phys.\ Lett.\  B {\bf 251}, 450 (1990);
G.~Rossi,
  Phys.\ Rev.\  D {\bf 29}, 852 (1984);
F.~Borzumati and G.~A.~Schuler,
  Z.\ Phys.\  C {\bf 58}, 139 (1993);
M.~Gl\"{u}ck, E.~Reya and M.~Stratmann,
  Phys.\ Rev.\  D {\bf 51}, 3220 (1995); Phys.\ Rev.\  D {\bf 54}, 5515 (1996)
  [arXiv:hep-ph/9605297];
G.~A.~Schuler and T.~Sjostrand,
  Z.\ Phys.\  C {\bf 68}, 607 (1995)
  [arXiv:hep-ph/9503384];
Phys.\ Lett.\  B {\bf 376}, 193 (1996)
  [arXiv:hep-ph/9601282];
 M.~Drees and R.~M.~Godbole,
  Phys.\ Rev.\  D {\bf 50}, 3124 (1994)
  [arXiv:hep-ph/9403229];
D.~de Florian, C.~Garcia Canal and R.~Sassot,
  Z.\ Phys.\  C {\bf 75}, 265 (1997)
  [arXiv:hep-ph/9608438];
M.~Klasen, G.~Kramer and B.~P\"{o}tter,
  Eur.\ Phys.\ J.\  C {\bf 1}, 261 (1998)
  [arXiv:hep-ph/9703302];
 G.~Kramer and B.~P\"{o}tter,
  Eur.\ Phys.\ J.\  C {\bf 5}, 665 (1998)
  [arXiv:hep-ph/9804352];
M.~Gl\"{u}ck, E.~Reya and I.~Schienbein,
  Phys.\ Rev.\  D {\bf 60}, 054019 (1999)
  [Erratum-ibid.\  D {\bf 62}, 019902 (2000)]
  [arXiv:hep-ph/9903337]; Phys.\ Rev.\  D {\bf 63}, 074008 (2001)
  [arXiv:hep-ph/0009348].

\bibitem{bussey}
  P.~J.~Bussey  [H1 Collaboration and ZEUS Collaboration],
  Nucl.\ Phys.\ Proc.\ Suppl.\  {\bf 126}, 17 (2004)
  [arXiv:hep-ex/0307071], {\it and references therein}.

\bibitem{Collins:1989gx}
  J.~C.~Collins, D.~E.~Soper and G.~Sterman,
  Adv.\ Ser.\ Direct.\ High Energy Phys.\  {\bf 5}, 1 (1988)
  [arXiv:hep-ph/0409313].

\bibitem{Braaten:2001sz}
  E.~Braaten and J.~Lee,
  Phys.\ Rev.\  D {\bf 65}, 034005 (2002)
  [arXiv:hep-ph/0102130].

\bibitem{Kroll:1955zu}
  N.~M.~Kroll and W.~Wada,
  Phys.\ Rev.\  {\bf 98}, 1355 (1955).

\bibitem{Faessler:1999de}
  A.~F\"{a}{\ss}ler, C.~Fuchs and M.~I.~Krivoruchenko,
  Phys.\ Rev.\  C {\bf 61}, 035206 (2000)
  [arXiv:nucl-th/9904024].

\bibitem{Qiu:2001ac}
  J.~W.~Qiu, R.~Rodriguez and X.~f.~Zhang,
  Phys.\ Lett.\  B {\bf 506}, 254 (2001)
  [arXiv:hep-ph/0102198].

\bibitem{CTEQ6}
  J.~Pumplin, D.~R.~Stump, J.~Huston, H.~L.~Lai, P.~Nadolsky and W.~K.~Tung,
  JHEP {\bf 0207}, 012 (2002)
  [arXiv:hep-ph/0201195].

\bibitem{KKP}
B.~A.~Kniehl, G.~Kramer and B.~P\"{o}tter,
Nucl.\ Phys.\  B {\bf 582}, 514 (2000) [arXiv:hep-ph/0010289].

\bibitem{BGK} E.~L.~Berger, L.~E.~Gordon and M.~Klasen, 
Phys.\ Rev.\  D {\bf 58}, 074012 (1998) [arXiv:hep-ph/9803387].

\bibitem{gv} {\it here we have used the Fortran code of:} 
L.~E.~Gordon and W.~Vogelsang, 
Phys.\ Rev.\  D {\bf 48}, 3136 (1993); 
Phys.\ Rev.\  D {\bf 50}, 1901 (1994).

\bibitem{arleo} F.~Arleo,
JHEP {\bf 0609}, 015 (2006) [arXiv:hep-ph/0601075].

\bibitem{iso_gamma}
  S.~Turbide, C.~Gale, E.~Frodermann and U.~Heinz,
  Phys.\ Rev.\  C {\bf 77}, 024909 (2008) [arXiv:0712.0732 [hep-ph]];
  I.~Vitev and B.~W.~Zhang,
  Phys.\ Lett.\  B {\bf 669}, 337 (2008)
  [arXiv:0804.3805 [hep-ph]]; 
  arXiv:0810.3194 [nucl-th].

\bibitem{saturation} {\it see, for example:}
  J.~Raufeisen, J.~C.~Peng and G.~C.~Nayak,
  Phys.\ Rev.\  D {\bf 66}, 034024 (2002)
  [arXiv:hep-ph/0204095];
  R.~Baier, A.~H.~Mueller and D.~Schiff,
  Nucl.\ Phys.\  A {\bf 741}, 358 (2004)
  [arXiv:hep-ph/0403201];
  F.~Gelis and J.~Jalilian-Marian,
  Phys.\ Rev.\  D {\bf 76}, 074015 (2007)
  [arXiv:hep-ph/0609066],
  {\it and references therein}.
    
\bibitem{Guo:1997it}
  X.~F.~Guo,
  Phys.\ Rev.\  D {\bf 58}, 036001 (1998)
  [arXiv:hep-ph/9711453];
  J.~W.~Qiu and X.~Zhang,
  Phys.\ Lett.\  B {\bf 525}, 265 (2002)
  [arXiv:hep-ph/0109210];
  Z.~B.~Kang and J.~W.~Qiu,
  Phys.\ Rev.\  D {\bf 77}, 114027 (2008)
  [arXiv:0802.2904 [hep-ph]].

\bibitem{parton_m_col}  {\it see, for example:}
  X.~F.~Guo and J.~W.~Qiu,
  Phys.\ Rev.\  D {\bf 53}, 6144 (1996)
  [arXiv:hep-ph/9512262];
  S.~Peigne,
  Phys.\ Rev.\  D {\bf 66}, 114011 (2002)
  [arXiv:hep-ph/0206138];
  R.~J.~Fries,
  Phys.\ Rev.\  D {\bf 68}, 074013 (2003)
  [arXiv:hep-ph/0209275];
  B.~Z.~Kopeliovich, J.~Raufeisen, A.~V.~Tarasov and M.~B.~Johnson,
  Phys.\ Rev.\  C {\bf 67}, 014903 (2003)
  [arXiv:hep-ph/0110221], {\it and references therein}.

\bibitem{thermal} {\it see, for example:}
S.~Turbide, C.~Gale, D.~K.~Srivastava and R.~J.~Fries,
  Phys.\ Rev.\  C {\bf 74}, 014903 (2006)
  [arXiv:hep-ph/0601042];
H.~van Hees and R.~Rapp,
  Nucl.\ Phys.\  A {\bf 806}, 339 (2008)
  [arXiv:0711.3444 [hep-ph]];
B.~K\"{a}mpfer and O.~P.~Pavlenko,
  Phys.\ Lett.\  B {\bf 391}, 185 (1997). {\it See also:}
N.~Armesto {\it et al.},
  J.\ Phys.\ G {\bf 35}, 054001 (2008)
  [arXiv:0711.0974 [hep-ph]];
F.~Arleo {\it et al.},
  arXiv:hep-ph/0311131;
P.~Aurenche,
  arXiv:hep-ph/0610218, {\it and references therein}.

\bibitem{EKS98}
  K.~J.~Eskola, V.~J.~Kolhinen and C.~A.~Salgado,
  Eur.\ Phys.\ J.\  C {\bf 9}, 61 (1999)
  [arXiv:hep-ph/9807297];
  K.~J.~Eskola, V.~J.~Kolhinen and P.~V.~Ruuskanen,
  Nucl.\ Phys.\  B {\bf 535}, 351 (1998)
  [arXiv:hep-ph/9802350].

\bibitem{fgs} L.~Frankfurt, V.~Guzey and M.~Strikman,
Phys.\ Rev.\  D {\bf 71}, 054001 (2005) [arXiv:hep-ph/0303022].

\bibitem{kumano} M.~Hirai, S.~Kumano and M.~Miyama,
Phys.\ Rev.\  D {\bf 64}, 034003 (2001) [arXiv:hep-ph/0103208];
M.~Hirai, S.~Kumano and T.~H.~Nagai,
Phys.\ Rev.\  C {\bf 76}, 065207 (2007) [arXiv:0709.3038 [hep-ph]].

\bibitem{dFS2003}
  D.~de Florian and R.~Sassot,
  Phys.\ Rev.\  D {\bf 69}, 074028 (2004)
  [arXiv:hep-ph/0311227].

\bibitem{EPS08}
  K.~J.~Eskola, H.~Paukkunen and C.~A.~Salgado,
  arXiv:0802.0139 [hep-ph].

\bibitem{impact}
R.~Vogt, Phys.\ Rev.\ C {\bf 70}, 064902 (2004); 
K.~J.~Eskola, Z.\ Phys.\ C {\bf 51},  633 (1991).

\bibitem{Peressounko:2006qs}
D.~Peressounko  [PHENIX Collaboration],
Nucl.\ Phys.\  A {\bf 783}, 577 (2007) [arXiv:hep-ex/0609037].

\bibitem{KleinBosing:2007bp} 
C.~Klein-Boesing  [PHENIX Collaboration],
J.\ Phys.\ G {\bf 35}, 044026 (2008) [arXiv:0710.2960 [nucl-ex]].

\bibitem{TAA}
  M.~L.~Miller, K.~Reygers, S.~J.~Sanders and P.~Steinberg,
  Ann.\ Rev.\ Nucl.\ Part.\ Sci.\  {\bf 57}, 205 (2007)
  [arXiv:nucl-ex/0701025];
S.~S.~Adler {\it et al.}  [PHENIX Collaboration],
  Phys.\ Rev.\ Lett.\  {\bf 91}, 072301 (2003)
  [arXiv:nucl-ex/0304022].

\end{thebibliography}
\end{document}